\documentclass[11pt,a4paper]{article}
\usepackage{epsfig}
\usepackage[T1]{fontenc}    
\usepackage{graphics}
\usepackage{graphicx}
\usepackage{pstricks,pst-coil,pst-fill,pst-plot}
\usepackage[fleqn]{amsmath}    
\usepackage{amssymb}    
\usepackage{amsfonts}   
\usepackage{verbatim}   
\usepackage{mathrsfs}   
\usepackage{dsfont}
\usepackage{euscript}
\usepackage{yfonts}
\usepackage{enumerate}     
\usepackage{amsthm}         
\usepackage{txfonts}
\usepackage{marvosym}
\usepackage{stmaryrd}
\usepackage{vmargin}        
\usepackage{wasysym}		

\usepackage{color}

\setmarginsrb{1.8cm}{2cm}{1.8cm}{2cm}{1cm}{1cm}{1cm}{1.6cm}
 \makeatletter
 \@addtoreset{equation}{section}
 \makeatother

\providecommand{\bysame}{\leavevmode\hbox to3em{\hrulefill}\thinspace}
\providecommand{\MR}{\relax\ifhmode\unskip\space\fi MR }

\providecommand{\href}[2]{#2}

\let\ua=\uparrow
\let\da=\downarrow
\let\tend=\rightarrow

\long\def\symbolfootnote[#1]#2{\begingroup%
\def\thefootnote{\fnsymbol{footnote}}\footnote[#1]{#2}\endgroup}

\newtheorem{theorem}{Theorem}[section]
\newtheorem{prop}[theorem]{Proposition}
\newtheorem*{theorem*}{Theorem}

\newtheorem{lemme}[theorem]{Lemma}

\def\Proof{\medskip\noindent {\it Proof --- \ }}

\def\qed{\hfill\rule{2mm}{2mm}}

\newcommand\beq{\begin{equation}}
\newcommand\enq{\end{equation}}
\newcommand\bem{\begin{multline}}
\newcommand\enm{\end{multline}}

\def\beqa{\begin{eqnarray}}
\def\eeqa{\end{eqnarray}}
\def\ba{\begin{array}}
\def\ea{\end{array}}

\newcommand{\f}[2]{{\ensuremath{%
    \mathchoice%
    {\dfrac{#1}{#2}}
    {\dfrac{#1}{#2}}
    {\frac{#1}{#2}}
    {\frac{#1}{#2}}
}}}
\newcommand{\tf}[2]{\ensuremath{#1/#2}}
\def\a{\alpha}

\def\be{\beta}

\def\Ga{\Gamma}

\def\de{\delta}

\def\eps{\epsilon}
\def\veps{\varepsilon}
\def\la{\lambda}

\def\sg{\sigma}
\def\vsg{\varsigma}

\def\ups{\upsilon}

\def\vth{\vartheta}

\def\om{\omega}
\def\vp{\varphi}

\newcommand{\mc}[1]{\ensuremath{\mathcal{#1}}}
\newcommand{\mf}[1]{\ensuremath{\mathfrak{#1}}}
\newcommand{\msc}[1]{\ensuremath{\mathscr{#1}}}

\newcommand{\bs}[1]{\ensuremath{\boldsymbol{#1}}}

\DeclareFontFamily{OT1}{pzc}{}
\DeclareFontShape{OT1}{pzc}{m}{it}{<-> s * [1.10] pzcmi7t}{}
\DeclareMathAlphabet{\mathpzc}{OT1}{pzc}{m}{it}

\def \i{ \mathrm i}

\newcommand{\ov}[1]{\ensuremath{\overline{#1}}}
\newcommand{\wt}[1]{\ensuremath{\widetilde{#1}}}
\newcommand{\wh}[1]{\ensuremath{\widehat{#1}}}

\newcommand{\Int}[2]{\ensuremath{\int\limits_{#1}^{#2}}}

\newcommand{\Fint}[2]{\ensuremath{\fint\limits_{#1}^{#2}}}

\newcommand{\sul}[2]{\ensuremath{\sum\limits_{#1}^{#2}}}
\newcommand{\pl}[2]{\ensuremath{\prod\limits_{#1}^{#2}}}

\newcommand{\R}{\ensuremath{\mathbb{R}}}
\newcommand{\Cx}{\ensuremath{\mathbb{C}}}

\newcommand{\Dp}[1]{\ensuremath{\partial_{#1}}}

\newcommand{\limit}[2]{\ensuremath{\underset{#1 \tend #2}{\longrightarrow} }}

\newcommand{\ex}[1]{\ensuremath{\e{e}^{#1}}}

\newcommand{\op}[1]{ \boldsymbol{ \texttt{#1} } }

\newcommand{\norm}[1]{\ensuremath{  || #1 || }}

\newcommand{\dd}{\mathrm{d}}
\newcommand{\e}[1]{\ensuremath{\mathrm{#1}}}

\newcommand{\intff}[2]{\ensuremath{ [  #1 \,; #2 ] }}
\newcommand{\intfo}[2]{\ensuremath{ [  #1 \,; #2 [ }}
\newcommand{\intof}[2]{\ensuremath{ ]  #1 \,; #2 ] }}
\newcommand{\intoo}[2]{\ensuremath{ ]  #1 \,; #2 [ }}

\begin{document}

\begin{center}
\begin{LARGE}
{\bf On the equilibrium measure for the Lukyanov integral}
\end{LARGE}

\vspace{1cm}

\vspace{4mm}

{\large Charlie Dworaczek Guera \footnote{e-mail: charlie.dworaczek@ens-lyon.fr}}%
\\[1ex]
Univ Lyon, ENS de Lyon, Univ Claude Bernard Lyon 1, CNRS, Unité de Mathématiques Pures et Appliquées, F-69342 Lyon, France \\[2.5ex]

{\large Karol K. Kozlowski \footnote{e-mail: karol.kozlowski@ens-lyon.fr}}%
\\[1ex]
Univ Lyon, ENS de Lyon, Univ Claude Bernard Lyon 1, CNRS, Laboratoire de Physique, F-69342 Lyon, France \\[2.5ex]

\par

\vspace{30pt}

\centerline{\bf Abstract} \vspace{1cm}
\parbox{12cm}{\small  In 2000, Lukyanov conjectured that a certain ratio of $N$-fold integrals should provide access, in the large-$N$ regime, to  the ground state expectation value of the exponential of the
Sinh-Gordon quantum field   in 1+1 dimensions and finite volume $R$.
This work aims at rigorously constructing the fundamental objects necessary to address the large-$N$ analysis of such integrals.
More precisely, we construct and establish the main properties of the the equilibrium measure minimising a certain $N$-dependent energy functional
that naturally arises in the study of the leading large-$N$ behaviour of the Lukyanov integral.
Our construction  allows us to heuristically advocate the leading term
in the large-$N$ asymptotic behaviour of the mentioned ratio of Lukyanov integrals, hence supporting Lukyanov's prediction -obtained by other means-
on the exponent $\sg$ of the power-law $N^{\sg}$ term of its asymptotic expansion as $N\tend + \infty$.}

\end{center}

\vspace{20pt}


\vspace{40pt}
\tableofcontents

\section{Introduction and statement of results}

\subsection{The separation of variable integral for the exponent of the field}

The seminal works of Al. Zamolodchikov \cite{ZalmolodchikovFirstIntroTBAForIQFT}, following the pioneering considerations of Yang-Yang \cite{YangYangNLSEThermodynamics}, introduced the concept of Thermodynamic
Bethe Ansatz (TBA) as a key tool allowing one to describe the ground state energies of integrable quantum field theories in finite volume. The
construction, as an input, utilises the model's $\op{S}$-matrix and describes the per-volume ground state energy in terms of a solution to
a non-linear integral equation. In the case of the Sinh-Gordon 1+1 dimensional quantum field theory, the TBA description was conjectured
simultaneously and independently by Al. Zamolodchikov \cite{ZalmolodchikovTBAForSinhGordon} and   Lukyanov \cite{LukyanovConjectureOfFieldExpValueSinhGAndRenormalization}.
In that case, there is a unique TBA equation which takes the form
\beq
\veps(\la) \, = \, 2 \mf{r} \sin\Big[ \tfrac{\pi }{ 1 + \mf{b}^2 } \Big]\cosh(\la) \, + \, \Int{\R}{} K(\la-\mu) \ln \Big[ 1\, + \, \ex{-\veps(\mu) }  \Big]
\quad \e{with} \quad  K(\la) \; = \;  \f{ 4 \cosh(\la)   \sin\Big[ \tfrac{\pi }{ 1 + \mf{b}^2 } \Big]  }{ \cosh(2\la) \, - \, \cos\Big[ \tfrac{ 2 \pi }{ 1 + \mf{b}^2 } \Big]  } \;.
\label{ecriture equation TBA pour Sinh Gordon}
\enq
This equation involves two parameters $\mf{r}, \mf{b}>0$. $\mf{r}=C m R$ with $C>0$ some constant, $R$ the model's volume and $m$
the mass parameter. Finally, $\mf{b}>0$ measures the interaction strength.

It was rigorously shown in \cite{FringKorffSchultzUVBehaviourofIQFT+ProofSolvabilityTBASinhG}
that, for any $\mf{r}>0$,  the non-linear integral equation \eqref{ecriture equation TBA pour Sinh Gordon} admits a unique solution in $L^{\infty}(\R)$.
With the solution at hand, the per-volume ground state energy admits the integral representation \cite{ZalmolodchikovTBAForSinhGordon}
\beq
-m \Int{ \R }{} \f{ \dd \la }{ 2\pi } \cosh(\la) \, \ln \Big[ 1\, + \, \ex{-\veps(\la) }  \Big] \;.
\enq
These TBA-like considerations were backed up by Bethe Ansatz calculations carried out for the lattice discretisation of the
finite-volume Sinh-Gordon quantum field theory in \cite{BytskoTeschnerSinhGordonFunctionalBA,TeschnerSpectrumSinhGFiniteVolume}.

\vspace{2mm}

By invoking an analogy with the classical method of separation of variables, Lukyanov \cite{LukyanovConjectureOfFieldExpValueSinhGAndRenormalization} conjectured
that the ground state expectation value of the exponential of the quantum Sinh-Gordon field $\ex{\a \vp}$ may be deduced from the data contained in the large-$N$ behaviour of the below integral
\beq
\mf{z}_N\big[V_{\a}] \; = \; \Int{ \R^N }{} \dd^N \la \pl{k<\ell}{N} \bigg\{ \sinh\Big[  (1+\mf{b}^2) (\la_k-\la_{\ell} )  \Big] \cdot \sinh\Big[  (1+\mf{b}^{-2})   (\la_k-\la_{\ell} )  \Big]    \bigg\}
\pl{k=1}{N} \ex{-  V_{\a}(\la_k) } \;.
\label{ecriture integrale de Lukyanov}
\enq
The potential $V_{\a}$ appearing above was expressed in terms of the solution $\veps$ to the TBA equation for the Sinh-Gordon model \eqref{ecriture equation TBA pour Sinh Gordon} as
\beq
V_{\a}(\la) \; = \; \mf{r} \cosh(\la) \, - \a  \la \, - \, \Int{\R}{} \f{ \dd \mu }{2\pi} \cdot \f{ \op{g}(\mu)  }{  \cosh(\la-\mu) }  \qquad \e{with} \qquad
\op{g}(\mu)\,= \, 2  \ln \Big[ 1\, + \, \ex{-\veps(\la) }  \Big] \;.
\enq

It was conjectured in \cite{LukyanovConjectureOfFieldExpValueSinhGAndRenormalization}  that, as $N \tend + \infty$,
\beq
\f{ \mf{z}_N\big[V_{\a(\mf{b}+\mf{b}^{-1})}]  }{ \mf{z}_N\big[V_{0}]  } \; = \; \bigg( \f{ N }{ \mf{r} } \bigg)^{ \f{ \a^2  }{2}  } \cdot \big< \ex{\a \vp} \big>_{\mf{r}} \cdot  \Big( 1+\e{o}(1)\Big) \;.
\label{ecriture prediction Lukyanov sur ratio fct partitions}
\enq
The constant term in these asymptotics $\big< \ex{\a \vp} \big>_{\mf{r}}$ was conjectured to coincide with the ground state expectation value of the exponent of the properly normalised Sinh-Gordon quantum
field in the finite volume $R$ theory. In this description, the fields were normalised so that, as $\mf{r} \tend + \infty$, the
two-point functions of the fields have trivial CFT-like normalisation in the short space-like  Minkowski-distance
regime, see \cite{LukyanovConjectureOfFieldExpValueSinhGAndRenormalization} for more details.

\vspace{2mm}

In fact,  Lukyanov's integral falls into a much larger class of $N$-fold integrals describing so-called form factors -\textit{i.e.} matrix elements- of local operators
in numerous quantum integrable models solvable by the quantum separation of variables method, see \textit{e.g.} \cite{BabelonQuantumInverseProblemConjClosedToda,DerkachovKorchemskyManashovXXXreelSoVandQopABAcomparaison,KozDerkachovManashovUnitaritySoVTransformModularXXZ,DerkachovManashovInterativeConstrEigenFctsSL2C,
GrosjeanMailletNiccoliFFofLatticeSineG,KazamaKomatsuNishimuraSoVLikeMIRepForPartFctRat6Vertex,KozIPForTodaAndDualEqns,KozUnitarityofSoVTransform,
NiccoliCompleteSpectrumAndSomeFormFactorsAntiPeriodicXXZ}. In such a setting the form factors are expressed as a ratio $\tf{ \msc{Z}[W_N] }{ \msc{Z}[V_N] }$
for certain potentials $V_N, W_N$ and where
\beq
 \msc{Z}[V_N] \, = \, \Int{ \msc{C}^N }{} \dd^N \la  \pl{k<\ell}{N} \bigg\{ \sinh\Big[  \pi \om_1 (\la_k - \la_{\ell})  \Big] \cdot \sinh\Big[ \pi \om_2 (\la_k - \la_{\ell})  \Big]    \bigg\}
\pl{k=1}{N} \ex{- V_N(\la_k) } \;.
\label{ecriture qSoV partition fct}
\enq
Above $\om_1, \om_2$ are related to a given model's coupling constants, and $\msc{C}$ is a model dependent curve in $\Cx$. The curve $\msc{C}$ may or may not be compact or closed.
Also, the potential $V_N$ may or may not depend on $N$. However, typically, it is \textit{not} varying with $N$ as $V_N(\la)=N U(\la)$ for some $N$ independent function $U$, \textit{i.e.}
the $N$-dependence, if present, is much more involved.
For such multiple integrals, one is usually interested in the $N\tend +\infty$ regime which allows one either to reach the thermodynamic or the continuum limit of the model.
Thus, on top of testing Lukyanov's conjecture, the possibility to study of the large-$N$ behaviour of this class of integrals will have numerous applications in the
field of quantum integrable models. We should mention that the above integral falls into the class of bi-orthogonal ensembles \cite{BorodinBiOrthogonalEnsIntroAnStudyOfExamples}.

\vspace{2mm}

It is clear that the class of quantum separation of variables integrals $ \msc{Z}[V_N] $ bears a strong structural ressemblance with the
spectral partition function of a random Hermitian matrix $M$ sampled from a measure $ \ex{-N \e{Tr}[V(M)]} \dd M$, with $\dd M$ being a properly normalised Lebesgue measure on the
statistically independent entries. Indeed, the latter takes the form
\beq
\msc{Z}_{N;\e{Herm}}[V] \; = \; \Int{ \R^N }{} \dd^N \la \pl{k<\ell}{N}  |\la_k - \la_{\ell}|^2 \pl{k=1}{N} \ex{-N V(\la_k) }
\label{ecriture fct partition random hermitian matrix}
\enq
In both cases \eqref{ecriture qSoV partition fct}-\eqref{ecriture fct partition random hermitian matrix},
there appears a one-body confining potential and a repulsive two-body interaction vanishing as the square of the spacing between the integration variables.
In fact, the two-body interaction is given by Vandermonde determinants in both cases: the square of a usual Vandermonde in the random matrix case \eqref{ecriture fct partition random hermitian matrix} and the
product of two-hyperbolic Vandermondes in the quantum separation of variables case \eqref{ecriture qSoV partition fct}. One could thus hope that the techniques allowing one to deal with
the large-$N$ behaviour of classical random matrix ensembles will also be fit for tackling the large-$N$ behaviour of quantum separation of variables issued integrals.
Unfortunately, the situation is way more intricate and takes its origin in crucial differences between these two types of integrals.
Genuinely, the potentials arising in the quantum separation of variables case are \textit{not}\symbolfootnote[2]{Else, indeed, the study of the large-$N$
behaviour of quantum separartion of variables issued multiple integrals would follow from the application of techniques developed \cite{KozBorotGuionnetLargeNBehMulIntMeanFieldTh}}
varying with $N$ as $V_N(\la)=N U(\la)$ for some $N$ independent function $U$. Thus, the two-body and the one-body interactions in $ \msc{Z}[V_N] $
evolve on different scales and one needs to dilatate the integration variables, in an appropriate fashion, so that both rescaled interactions
equilibrate. While in the random matrix case the two-body interaction was behaving trivially under rescalings, this is not anymore the case in the
quantum separation of variables setting. This introduces several additional scales in $N$ to the problem what makes numerous of the steps
developed for the random matrix case very tricky, technically speaking, to set in. We would like to mention that, in fact, certain instances of integrals of the type
\eqref{ecriture qSoV partition fct} did in fact appear directly in the random matrix literature.
More precisely, the spectral part of a random Hermitian matrix's sampled from a measure $\propto \ex{-N \e{Tr}[V(M)-AM]} \dd M$ with $A=\e{diag}(a_1,\dots, a_N)$,
$a_k=\tf{(k-1)}{N}$, admits the integral representation \cite{ClaeysWangIntroRHPForBioRthSinhLikePlys}
\beq
\msc{Z}_{N;\e{Source}}[V] \; = \; \Int{ \R^N }{} \dd^N \la \pl{k<\ell}{N}  \Big\{ (\la_k - \la_{\ell})\sinh\big(\la_k-\la_{\ell}\big) \Big\} \pl{k=1}{N} \ex{-N V(\la_k) } \;.
\label{ecriture fct partition random hermitian matrix ext source}
\enq
The asymptotic expansion of such integrals may be deal with by using the techniques developed in \cite{KozBorotGuionnetLargeNBehMulIntMeanFieldTh}.
The work \cite{ClaeysWangIntroRHPForBioRthSinhLikePlys} proposed a Riemann--Hilbert approach for bi-orthogonal polynomials
that could, in principle, allow one to extract the large-$N$ behaviour of \eqref{ecriture fct partition random hermitian matrix ext source} and, more generally,
\eqref{ecriture qSoV partition fct}. However, in the case of the quantum separation of variables issued integral,
the necessity for rescaling the integration variables would introduce numerous technical complications to the large-$N$
analysis technique developed in \cite{ClaeysWangIntroRHPForBioRthSinhLikePlys}. In particular, it would demand to
have a highly detailed control on the $N$-dependent equilibrium measure that will be obtained in the present paper.

The first progress in this direction of achieving a large-$N$ analysis of multiple integrals of the form \eqref{ecriture qSoV partition fct}
was achieved in \cite{KozBorotGuionnetLargeNBehMulIntOfToyModelSoVType} where techniques allowing one to deal with $N$-dependent two-body interactions were developed.
The aim of this work is to push further the results obtained in  \cite{KozBorotGuionnetLargeNBehMulIntOfToyModelSoVType} and lay the ground  for rigorously justifying
the presence of the the power-law term in $N$ in \eqref{ecriture prediction Lukyanov sur ratio fct partitions}, and, in a second stage, for obtaining rigorously the
whole expansion up to $\e{o}(1)$. In order to apply concentration of measure techniques
which were first developed for $\be$-ensembles in \cite{BenArousGuionnetLargeDeviationForWignerLawLEadingAsymptOneMatrixIntegral} and, later, extended so as to
allow to deal with more complex integrals in \cite{BorotGuionnetAsymptExpBetaEnsMultiCutRegime,KozBorotGuionnetLargeNBehMulIntMeanFieldTh,KozBorotGuionnetLargeNBehMulIntOfToyModelSoVType},
one first needs to have a good grasp on the so-called equilibrium measure. As explained in  \cite{KozBorotGuionnetLargeNBehMulIntOfToyModelSoVType},
in the case of the partition function \eqref{ecriture integrale de Lukyanov}, the latter corresponds to the unique minimiser of an $N$-dependent functional on $\mc{M}^{1}(\R)$, the space of probability measures on $\R$.
The construction of the equilibrium measure is the main achievement of this work. As mentioned, this may also pave the way to the Riemann--Hilbert
analysis of the large-$N$ behaviour of \eqref{ecriture integrale de Lukyanov} by means of bi-orthogonal polynomials.
Our result allows us to back up the prediction on the leading large-$N$ behaviour given in \eqref{ecriture prediction Lukyanov sur ratio fct partitions},
although the lack of sharp estimates on the remainders does not allow us to turn our findings into a rigorous proof.

\vspace{3mm}

The paper is organised as follows.
In Subsection \ref{SousSection Resultats principaux} we gather the main results obtained in this work.
Section \ref{Section N dependent equilibrium measure} is devoted to the construction of the equilibrium measure
governing the leading large-$N$ behaviour of the Lukyanov integral. In Subsection \ref{SousSection pts generale mesure eq},
we establish basic properties of the equilibrium measure. In Subsection \ref{SousSection operateur WH}, we recall
the explicit representation for the inverse of a truncated Wiener-Hopf operator that arises in the characterisation of the
equilibrium measure's density. In Subsection \ref{SousSection Rep Int inverse sur V prime}
we establish a convenient integral representation for the equilibrium measure's density. Finally, in Subsection \ref{SousSection support mesure eq},
we construct its support. Section \ref{Section DA integrale interpolante} is devoted to obtaining the first few terms in the large-$N$ asymptotic expansion of a certain integral
\textit{versus} the equilibrium measure that plays a role in the problem of our interest.

\subsection{The main results}
\label{SousSection Resultats principaux}

It is easy to see that the repulsive nature of the $\sinh$ two-body interaction and the confining nature of the potential are of the same order of magnitude in $N$
on a scale $\ln N$. Hence, so as to deal with finite quantities, it appears convenient to rescale the integration variables in \eqref{ecriture integrale de Lukyanov} by $\ln N$. Then, it holds
$\mf{z}_N\big[V_{\a}] \; = \; \big[ \ln N \big]^N  \mc{Z}_N\big[V_{N;\a}] $ with
\beq
\mc{Z}_N\big[V_{N;\a}] \; = \; \Int{ \R^N }{} \dd^N \la \pl{a<b}{N} \bigg\{ \sinh\Big[ \f{ \ov{\om}_1}{2} (\la_a - \la_b)  \Big] \cdot \sinh\Big[ \f{ \ov{\om}_2}{2} (\la_a - \la_b)  \Big]    \bigg\}
\pl{a=1}{N} \ex{- N \tau_N V_{N;\a}(\la_a) } \;.
\enq
 The two periods $\ov{\om}_a$ grow with $N$ as
\beq
\ov{\om}_a \; = \; 2\pi \tau_N \om_a \qquad \e{with} \qquad \tau_N \, = \,  \ln N  \;.
\label{definition periode omega a bar}
\enq
The rescaled confining potential takes the form
\beq
V_{N;\a}(\la) \, = \, \f{ \mf{r} }{ N \tau_N} \cosh\big[ \tau_N \la \big]  \, - \, \f{ \a \la   }{ N }
\, - \, \Int{\R}{} \f{ \dd \mu }{2\pi N   } \cdot \f{ \op{g}( \tau_N \mu)  }{  \cosh\big[ \tau_N (\la-\mu) \big] } \;.
\label{introduction potientiel V N alpha}
\enq
The connection to the Lukyanov integral imposes the following from for the periods
\beq
\om_1 \, = \, \f{ 1 + \mf{b}^2 }{ \pi } \quad \e{and} \quad \om_2 = \f{ 1 + \mf{b}^{-2} }{\pi} \,.
\label{ecriture spécification valeur periodes pour integrale de Lukyanov}
\enq

Following the techniques of \cite{BenArousGuionnetLargeDeviationForWignerLawLEadingAsymptOneMatrixIntegral}
and their adaptation to the $N$-dependent setting developed in \cite{KozBorotGuionnetLargeNBehMulIntOfToyModelSoVType}, one may show that
\beq
\mc{Z}_N\big[V_{N;\a}] \; = \; \exp\bigg\{- N^2 \tau_N  \underset{\mu \in \mc{M}^1(\R)}{\e{inf}} \Big\{ \mc{E}_N[\mu]  \Big\}  \, + \, \e{O}\big( N \tau_N^2 \big) \bigg\} \;,
\enq
in which the $N$-dependent functional on  $\mc{M}^1(\R)$ takes the form
\beq
\mc{E}_N[\mu] \;  = \; \Int{ \R }{} \dd \mu(s) V_{N;\a}(s) \, - \, \f{1}{2\tau_N} \Int{}{} \dd\mu(s) \dd \mu(t) \ln \Bigg\{ \pl{a=1}{2} \sinh\Big( \ov{\om}_a \f{s-t}{2} \Big) \Bigg\} \;.
\label{definition fnelle energie N dpdte}
\enq
$\mc{E}_N$ is strictly convex, lower-continuous and has compact level sets, see \cite{KozBorotGuionnetLargeNBehMulIntOfToyModelSoVType} for more details.
As such, it admits a unique minimiser on $\mc{M}^1(\R)$ denoted by $\wh{\mu}_{\e{eq};\a}$. The knowledge of this minimum thus allows one to access to
the first few terms in the asymptotic expansion of $\ln \mc{Z}_N\big[V_{N;\a}]$.

Our main result is gathered in the

\begin{theorem}
\label{Theoreme principal}

The equilibrium measure $\wh{\mu}_{\e{eq};\a}$. is Lebesgue continuous with a density $\wh{\varrho}_{\e{eq};\a}$ given by the square root of an analytic function.
There exists $\mf{r}_0$ such that, for any $N$ and $\mf{r} \geq \mf{r}_0$, it is supported on the segment $\sg_{N;\a}=\intff{ a_{N;\a} }{ b_{N;\a} }$. Moreover,
there exists $N_0$ such that, for any $N \geq N_0$ it is given by
\beq
\wh{\varrho}_{\e{eq};\a} \, = \, \mc{W}_N\big[ V^{\prime}_{N;\a} \bs{1}_{\sg_{N;\a}}   \big]_{\mid \substack{ a_N \hookrightarrow a_{N;\a} \\  b_N \hookrightarrow b_{N;\a}  }   }\;.
\enq
where $\mc{W}_N$ is the integral transform given by \eqref{ecriture transformation integrale WN}
in which one should specify the periods $\om_a$ as in \eqref{ecriture spécification valeur periodes pour integrale de Lukyanov}.

The endpoints $a_{N;\a}, b_{N;\a} $ of the support admit the large-$N$ expansion
\bem
\tau_N b_{N;\a} \,  = \,  \ln \bigg( \f{ \mf{d}_0 N }{ 2 } \bigg) \, + \, \f{ \a }{  N (1+\mf{b}^2)(1+\mf{b}^{-2})  }  \\ \, + \,
\f{1}{N^2} \Bigg\{ \mf{d}_1 \bigg(  1 \, + \, \f{2 \mc{F}[\op{g}](\i) }{ \pi \mf{r} } \bs{1}_{1<\zeta} \Big( 1+\tfrac{\a}{\pi} \Big) \bigg)
\, - \,  \f{\a^2 }{2 (1+\mf{b}^2)^2(1+\mf{b}^{-2})^2  } \Bigg\} \, + \, \e{O}\bigg(  \f{1}{N^{3-\eta}  } \bigg)
\end{multline}
and
\bem
\tau_N a_{N;\a} \,  = \,  -\ln \bigg( \f{ \mf{d}_0 N }{ 2 } \bigg) \, + \, \f{ \a }{  N (1+\mf{b}^2)(1+\mf{b}^{-2})  }   \\
\, - \, \f{1}{N^2} \Bigg\{  \mf{d}_1 \bigg(  1 \, + \, \f{2 \mc{F}[ \op{g} ](\i) }{ \pi \mf{r} } \bs{1}_{1<\zeta} \Big( 1+\tfrac{\a}{\pi} \Big) \bigg)
\, + \,  \f{\a^2 }{2  (1+\mf{b}^2)^2(1+\mf{b}^{-2})^2  } \Bigg\} \, + \, \e{O}\bigg( \f{1}{N^{ 3 -\eta } }\bigg)
\end{multline}
where $\eta>0$ is fixed but can be taken as small as need be. The two constants $\mf{d}_0, \mf{d}_1$ arising in this expansion take the form
\beq
\mf{d}_0 \, = \, \f{ 2 }{ \mf{r} \sqrt{\pi} } \pl{\ups = \pm }{} \bigg\{  \big( 1+\mf{b}^{2\ups} \big)^{ \f{-1}{ 2 ( 1+\mf{b}^{2\ups} )  } } \cdot \Ga \bigg( \f{1}{ 2 ( 1+\mf{b}^{2\ups} )  } \bigg) \bigg\}
\quad and \quad
\mf{d}_1 \, = \, \f{ \mf{r}^2 }{ \pi}  \pl{\ups = \pm }{} \bigg\{  \f{ \sin\Big[ \tfrac{\pi}{ 2 ( 1+\mf{b}^{2\ups} )} \Big]  }{ 1+\mf{b}^{2\ups} }  \bigg\} \;.
\enq

\end{theorem}

The hardest part of the theorem consists in proving the form of the endpoints, the rest of its content follows from the techniques already developed in
\cite{KozBorotGuionnetLargeNBehMulIntOfToyModelSoVType}.

\vspace{3mm}

One way to obtain the large-$N$ behaviour of the ratio: $ \tf{ \mf{z}_N\big[V_{\a}] }{ \mf{z}_N\big[V_{0}]  }$, \textit{i.e.} $\tf{ \mc{Z}_N\big[V_{N;\a}] }{ \mc{Z}_N\big[V_{N;0}]  }$
is to start from the differential identity
\beq
\Dp{\a} \ln \mc{Z}_N\big[V_{N;\a}] \; = \; N \tau_N \mathbb{E}_{N;\a}\bigg[  \Int{}{} \xi \dd \op{L}_N^{(\bs{\la}_N)}(\xi)  \bigg] \qquad \e{with} \qquad
\op{L}_N^{(\bs{\la}_N)}\, = \, \f{1}{N} \sul{a=1}{N} \de_{\la_a}
\enq
being the empirical distribution of the integration variables, $\bs{\la}_N = \big( \la_1, \dots, \la_N \big)$, and $\mathbb{E}_{N;\a}$ referring to the expectation in respect to the probability measure
$\mathbb{P}_{N;\a}$ on $\R^N$ with density
\beq
\mf{p}_{N;\a}\big(\bs{\la}_N\big) \; = \; \f{ 1 }{ \mc{Z}_N\big[V_{N;\a}]  }  \pl{a<b}{N} \bigg\{ \sinh\Big[ \f{ \ov{\om}_1}{2} (\la_a - \la_b)  \Big] \cdot \sinh\Big[ \f{ \ov{\om}_2}{2} (\la_a - \la_b)  \Big]    \bigg\}
\pl{a=1}{N} \ex{- N \tau_N V_{N;\a}(\la_a) } \;.
\enq
Using concentration of measure techniques as in \cite{KozBorotGuionnetLargeNBehMulIntOfToyModelSoVType} one may show that under $\mathbb{P}_{N;\a}$
the empirical measure concentrates around the equilibrium measure in the sense that for smooth functions growing at most polynomially at
infinity it holds
\beq
\bigg| \mathbb{E}_{N;\a}\bigg[  \Int{}{} \phi(\xi) \dd \Big( \wh{\mu}_{\e{eq};\a}-\op{L}_N^{(\bs{\la}_N)}\Big)(\xi)  \bigg]  \bigg| \, \leq \, C \f{\tau_N }{ \sqrt{ N} } \;.
\enq
for some constant depending on $\phi$. This reasoning then entails that
\beq
\Dp{\a} \ln \mc{Z}_N\big[V_{N;\a}] \; = \; N \tau_N    \Int{}{} \xi \dd  \wh{\mu}_{\e{eq};\a}(\xi)   \, + \, \e{O}\Big( \sqrt{ N} \tau_N^2  \Big) \;.
\enq
The remainder is uniform in $\a$.
It is important to stress that the bounds issuing from the concentration of measure estimates are only \textit{a priori} bounds. These may be improved by using the machinery of loop equations which
allow one to improve  the rigidity of the fluctuations, see \cite{BorotGuionnetAsymptExpBetaEnsOneCutRegime} for an implementation of the method in the case of
$\be$-ensembles, and compute the subdominant corrections contained in the $\e{O}$
remainder above. However, in the case of the present multiple integral, there arise several technical difficulties in the analysis of the associated system of loop equations
which go beyond the scope of the present analysis. We would however like to point out that under the specialisation \eqref{ecriture spécification valeur periodes pour integrale de Lukyanov}, it holds
\beq
N \tau_N \Int{}{} \xi \dd  \wh{\mu}_{\e{eq};\a}(\xi) \, = \, \f{ \a  \ln N }{ (1+\mf{b}^2)(1+\mf{b}^{-2}) } \, + \, \e{O}\big(1\big) \;,
\enq
with a remainder that is uniform in $\a$. See Proposition \ref{Proposition interpolating integral} for the detailed statement.

This leads to the suggestive results
\beq
\ln \bigg( \f{ \mc{Z}_N\big[V_{N;\a(\mf{b}+\mf{b}^{-1})^2}] }{ \mc{Z}_N\big[V_{N;0}] } \bigg)\; = \; \f{\a^2 }{2} \ln N  \, + \, \mf{R}_N
\enq
with $\mf{R}_N$ a remainder that we are able only to estimate as $ \e{O}\Big( \sqrt{ N} \tau_N^2  \Big)$. However, we expect that this
is an overestimate and that eventually, the machinery of loop equations will allow us to prove that $\mf{R}_N=\e{O}(1)$. Thus, while not being a rigorous proof
thereof, the above provides a strong check of Lukyanov's conjecture.





\section{The $N$-dependent equilibrium measure}
\label{Section N dependent equilibrium measure}

From now on, we keep $\om_1, \om_2$ arbitrary and shall specify our results to the setting \eqref{ecriture spécification valeur periodes pour integrale de Lukyanov} only at the very end.

\subsection{General properties of the equilibrium measure}
\label{SousSection pts generale mesure eq}

Consider the multiple integral
\beq
\mf{X}_M \, = \, \Int{\R^M}{} \dd^M \la \pl{a<b}{M} \bigg\{ \sinh\Big[ \pi \om_1 \tau_N (\la_a - \la_b)  \Big] \cdot \sinh\Big[ \pi \om_2 \tau_N  (\la_a - \la_b)  \Big]    \bigg\}^{\f{1}{\tau_N}}
\pl{a=1}{N} \ex{- M V_{N;\a}(\la_a) } \;.
\enq
There $N$ is fixed and to be considered as an outer parameter. This kind of integral has been studied in \cite{KozBorotGuionnetLargeNBehMulIntOfToyModelSoVType}. It was shown there that
\beq
\lim_{M\tend + \infty} \Big\{ \f{1}{M^2} \ln \mf{X}_M \Big\} \, = \, - \underset{\mu \in \mc{M}^1(\R)}{\e{inf}}  \mc{E}_N[\mu]
\enq
with $ \mc{E}_N$ as defined in \eqref{definition fnelle energie N dpdte}. The minimum is attained at a unique measure $\wh{\mu}_{\e{eq};\a}$ that has compact support given by a
finite union of segments and
is Lebesgue continuous with a density $\wh{\varrho}_{\e{eq};\a}$ given by the square root of an function analytic in an open neighbourhood of the support\symbolfootnote[2]{In our $N$ dependent setting,
the size of this neighbourhood will naturally depend on $N$}. In particular, the density is smooth in the interior of the support and vanishes at least as a square root
at the edges of the support.

\begin{lemme}
 \label{Lemme connexite du support de mu eq}
There exists $\mf{r}_0 > 0$ such that, for any $\mf{r}\geq \mf{r}_0$, the equilibrium measure has connected support
\beq
\sg_{N;\a} \; = \; \e{supp}\big[\, \wh{\mu}_{\e{eq};\a} \big] \, = \, \intff{ a_{N;\a} }{ b_{N;\a} } \;.
\label{ecriture support mesure equilibre}
\enq
\end{lemme}

\Proof

It is a classical fact that the support of the equilibrium measure will be connected, \textit{viz}. of the
form \eqref{ecriture support mesure equilibre}, as soon as $V_{N;\a}$ is strictly convex, see \textit{e.g.}
Appendix C of \cite{KozBorotGuionnetLargeNBehMulIntOfToyModelSoVType}

A direct calculation starting from \eqref{introduction potientiel V N alpha} yields
\beq
V_{N;\a}^{\prime\prime}(\la) \, = \, \f{ \mf{r}\tau_N  }{ N } \cosh\big[ \tau_N \la \big]
\, - \, \tau_N^2  \Int{\R}{} \f{ \dd \mu }{2\pi N   } \cdot \Bigg\{ \f{ 1  }{  \cosh\big[ \tau_N (\la-\mu) \big] }
\, - \, 2 \f{   \sinh^2\big[ \tau_N (\la-\mu) \big] }{  \cosh^3 \big[ \tau_N (\la-\mu) \big] }  \Bigg\} \cdot \op{g}( \tau_N \mu) \;.
\enq
It is easy to infer from the non-linear integral equation satisfied by $\veps$, that there exist $\mf{r}$-independent $c_{\veps}, c^{\prime}_{\veps}>0$ such that
\beq
\ex{- \veps(\la)  } \; \leq \; c_{\veps} \ex{ - \mf{r} c^{\prime}_{\veps} \cosh(\la) } \;.
\enq
One thus gets the lower bound
\bem
V_{N;\a}^{\prime\prime}(\la) \, \geq \, \f{ \mf{r}\tau_N  }{ N } \cosh\big[ \tau_N \la \big] \, - \,  c_{\veps} \tau_N^2 \ex{ - \mf{r} c^{\prime}_{\veps}  }
 \Int{\R}{} \f{ \dd \mu }{2\pi N   } \cdot \Bigg| \f{ 1  }{  \cosh\big[ \tau_N \mu \big] }
\, - \, 2 \f{   \sinh^2\big[ \tau_N \mu \big] }{  \cosh^3 \big[ \tau_N \mu \big] }  \Bigg| \\
\, \geq \, \f{ \tau_N  }{ N } \Big\{ \mf{r} \, -\, C \ex{ - \mf{r} c^{\prime}_{\veps}  } \Big\}\, >\, 0
\end{multline}
where the last bound follows provided that $\mf{r}\geq \mf{r}_0$ for some $\mf{r}_0>0$. \qed

\vspace{2mm}

It is a standard fact, see \textit{e.g.} \cite{DeiftOrthPlyAndRandomMatrixRHP} that $\wh{\mu}_{\e{eq};\a}$ corresponds to the unique solution to the variational problem
\beq
 V_{N;\a}(\la) \, - \, \f{1}{\tau_N} \Int{}{} \hspace{-1mm} \dd \, \wh{\mu}_{\e{eq};\a}(s) \ln \Bigg[  \pl{a=1}{2} \sinh\Big( \ov{\om}_a \f{\la-s}{2} \Big) \Bigg]
\qquad  \left\{ \ba{cc} = \;  C_{\e{eq};\a}^{(N)} & \la \in \sg_{N;\a}  \vspace{2mm}\\
                      > \;  C_{\e{eq};\a}^{(N)} & \la \in \R \setminus \sg_{N;\a}
\ea  \right.\; .
\label{ecriture equation variationnelle pour le potentiel effectif}
\enq
First of all, due to the smoothness of the equilibrium's measure density and its square root vanishing at the edges, we could
formulate this problem in the strong sense, \textit{i.e.} pointwise, and not \textit{a.e.}. Also, we stress that
due to the strict convexity of $V_{N;\a}$ for $\mf{r}\geq \mf{r}_0$, the second condition is immediately satisfied,
see Appendix C of \cite{KozBorotGuionnetLargeNBehMulIntOfToyModelSoVType} for more details.

This variational characterisation of the equilibrium measure allows one to obtain \textit{a priori} upper/lower bounds on the endpoints $a_{N;\a}/b_{N;\a}$.

\begin{lemme}
\label{Lemme borne inf sup sur tailler du support de la mesure equilibre}

 There exists $N_{0}>0$ and $\vsg>0$ such that, for any $N \geq N_0$,
\beq
a_{N;\a} \, \leq \, - \vsg \qquad and \qquad b_{N;\a} \, \geq \, \vsg \;.
\enq

\end{lemme}

\Proof

The proof goes by contradiction. Thus assume that for any $\vsg>0$ and  $N_0$ there exists $N\geq N_0$ such that
\beq
a_{N;\a} \, \geq \, - \vsg \qquad \e{or} \qquad b_{N;\a} \, \leq \, \vsg \;.
\enq
One may take $\vsg < \tf{1}{4}$ and extracting sub-sequences if need be, one gets that there exists a sequence $N_k\tend +\infty$ such that, for any $k$, $b_{N_k;\a} \, \leq \, \vsg$
or, for any $k$, $a_{N_k;\a} \, \geq \, -\vsg$. We discuss in detail the first case scenario, the second one
can be excluded in the same fashion.

We start by introducing the so-called effective potential
\beq
V_{\e{eff}}(\la) \; = \;  V_{N;\a}(\la) \, + \, \Int{}{} \hspace{-1mm} \dd \, \wh{\mu}_{\e{eq};\a}(s)\mf{f}_N(\la,s)  \qquad \e{with} \qquad
\mf{f}_N(\la,s) \, = \, \f{- 1}{\tau_N}  \ln \bigg[  \pl{a=1}{2} \sinh\Big( \ov{\om}_a \tfrac{\la-s}{2} \Big) \bigg]  \;.
\enq
One has that
\beq
\Dp{\la} \mf{f}_N(\la,s) \, = \,- \sul{a=1}{2} \pi \om_a\coth\big[ \pi \om_a \tau_N(\la-s) \big] \, < \, 0 \quad \e{for} \quad \la \geq s \;,
\enq
which entails that $\mf{f}_N(\la,s) - \mf{f}_N(\mu,s) <0$ for any $\la> \mu \geq s$. One may thus decompose, for any $\tf{1}{2}> \la > \tf{1}{4}$,
\bem
V_{\e{eff}}( \la ) \, - \, V_{\e{eff}}\big( b_{N_k;\a} \big) \, = \,  V_{N_k;\a}(\la) \, - \,  V_{N_k;\a}\big( b_{N_k;\a} \big)
\, + \, \Int{}{} \hspace{-1mm} \dd \, \wh{\mu}_{\e{eq};\a}(s)\Big( \mf{f}_{N_k}\big( \la, s \big) \, -\, \mf{f}_{N_k}\big( \tfrac{1}{4},s \big) \Big) \\
\, + \, \Int{}{} \hspace{-1mm} \dd \, \wh{\mu}_{\e{eq};\a}(s)\Big(\mf{f}_{N_k}\big( \tfrac{1}{4},s \big) \, -\, \mf{f}_{N_k}\big( b_{N_k;\a},s \big) \Big) \;.
\label{ecriture ecriture difference potentiels effectifs}
\end{multline}
One may then bound each term as follows. Since $\tf{1}{4}> \vsg \geq  b_{N_k;\a}$, the last line produces a purely negative contribution. Further,
it is direct to estimate that in the range of $\la$s considered $ V_{N_k;\a}(\la) \, = \, \e{o}(1)$ as $k \tend + \infty$. Likewise, if $-\tf{3}{4} \leq b_{N_k;\a} \leq \vsg $
then one has that $V_{N_k;\a}\big( b_{N_k;\a} \big) \, = \, \e{o}(1)$, while for  $b_{N_k;\a} \leq -\tf{3}{4}$ it holds for $k$ large enough that
$V_{N_k;\a}\big( b_{N_k;\a} \big) \geq  V_{N_k;\a}(\la)$. Thus, whatever the regime, it holds
\beq
V_{N_k;\a}\big( b_{N_k;\a} \big) \, - \,   V_{N_k;\a}(\la) \, \leq \, \e{o}(1)  \;.
\enq
Finally, one has, for $\la > s $ that
\beq
\mf{f}_N(\la,s) \, = \, - \pi (\om_1+\om_2)(\la-s) \, + \, \e{O}\Big( \f{1}{\tau_N} \sul{a=1}{2} \ex{-\ov{\om}_a(\la-s) } \Big) \;.
\enq
Thus, since $\la$ and $\tf{1}{4}$ are uniformly away from $s$ in the integral arising in the first line of \eqref{ecriture ecriture difference potentiels effectifs}, one gets that
\beq
V_{\e{eff}}( \la ) \, - \, C_{\e{eq};\a}^{(N)}  \, = \, V_{\e{eff}}( \la ) \, - \, V_{\e{eff}}\big( b_{N_k;\a} \big) \,  \leq  \,\e{o}(1) \, - \,  \pi (\om_1+\om_2)\big( \la - \tfrac{1}{4} \big) \, < \, 0  \;.
\enq
Since $\la \not \in \sg_{N;\a}$, this contradicts the variational equation \eqref{ecriture equation variationnelle pour le potentiel effectif}. \qed

One knows from Lemma \ref{Lemme connexite du support de mu eq} that the equilibrium measure is supported on the segment $\intff{ a_{N;\a} }{ b_{N;\a} }$.
Since $\wh{\varrho}_{\e{eq};\a}$ is smooth on $\intoo{ a_{N;\a} }{ b_{N;\a} }$ and admits at worst square root singularities
at the edges $a_{N;\a}, b_{N;\a}$, one has that $\wh{\varrho}_{\e{eq};\a}\in  H_{s}( \intff{ a_{N;\a} }{ b_{N;\a} } )$
for any $0<s < \tf{1}{2}$. In particular, one may differentiate the first relation given in \eqref{ecriture equation variationnelle pour le potentiel effectif}
what yields the singular-integral equation satisfied by the equilibrium measure's density
\beq
\sul{a=1}{2}\pi \om_a \Fint{ a_{N;\a}  }{ b_{N;\a} } \hspace{-1mm} \dd s \, \wh{\varrho}_{\e{eq};\a}(s) \coth\Big[ \ov{\om}_a\tfrac{ \la-s }{ 2 } \Big] \; = \;  V_{N;\a}^{\prime}(\la)  \;.
\label{ecriture eqn int sing pour densite}
\enq
Now, Lemma \ref{Lemme borne inf sup sur tailler du support de la mesure equilibre} ensures that $b_{N;\a}-a_{N;\a}\, \geq \, 2\vsg>0$, so that $\wh{\varrho}_{\e{eq};\a}$ solves a truncated Wiener-Hopf equation
in which the renormalised difference of boundaries satisfies
\beq
\tau_N b_{N;\a}\, -\, \tau_N a_{N;\a} \limit{N}{+ \infty} + \infty \;.
\enq
This allows one to invoke Riemann--Hilbert techniques so as to solve the equation in the large-$N$ regime and
will provide the starting starting point for characterising the measure and its support thoroughly in the large-$N$ regime.

\subsection{A truncated Wiener-Hopf based representation for the equilibrium measure}
\label{SousSection operateur WH}

In the following, we denote by $H_s(\R)$ the $s^{\e{th}}$ Sobolev space, \textit{viz}.
\beq
H_s(\R) \, = \, \bigg\{ f \in L^2(\R) \, : \, \Int{\R}{}\dd k \big|\mc{F}[f](k)\big|^2(1+k^2)^{s}  \, < \, + \infty\bigg\} \;.
\enq
Here, the Fourier transform $\mc{F}$ is defined, for $g \in L^1(\R)$, as
\beq
\mc{F}[g](\mu) \; =   \Int{ \R }{} \!  \dd \eta \, g(\eta) \ex{\i \mu \eta } \;.
\enq
Then, for any closed $F \subset \R$, $H_s(F)$ is the space of functions $f \in H_s(\R)$ such that $\e{supp}[f]\subset F$.

It was established in \cite{KozBorotGuionnetLargeNBehMulIntOfToyModelSoVType} that, provided that $b_N-a_N\, >\, \eta$ for some $\eta>0$
and that $N$ is large enough, the singular integral operator $\mc{S}_N \, : \, H_{s}( \intff{a_N}{b_N} ) \tend H_{s}( \R )$
with $0< s < \tf{1}{2}$, defined for sufficiently regular functions $h$ by the integral transform
\beq
\mc{S}_N[h](\la) \; = \; \sul{a=1}{2}\pi \om_a \Fint{}{} \hspace{-1mm} \dd \mu \, h(\mu) \coth\Big[ \ov{\om}_a\tfrac{ \la-\mu }{ 2 } \Big]
\enq
is invertible on the co-dimension $1$ subspace of $H_{s}( \R )$
\beq
\mf{X}_s(\R) \, = \, \bigg\{ g \in H_s(\R) \, : \, \mc{J}\big[ g \big] \; = \; 0 \Big\}
\qquad \e{with} \qquad
  \mc{J}\big[ g \big] \; =  \hspace{-1mm} \Int{\R+\i\eps^{\prime}}{} \hspace{-1mm} \f{\dd \mu }{2\i\pi} \, \chi_{11}(\mu)
\ex{-\i \ov{b}_{N} \mu} \mc{F}\big[ g \big]( \tau_N \mu) \;.
\label{ecriture fnelle contraintes}
\enq
Above and in the following, $\eps^{\prime}>0$ is arbitrary but taken sufficiently small and we agree upon
\beq
x_{N} \, = \, b_{N} - a_{N}\; , \quad \ov{x}_{N} \, = \, \tau_N x_{N} \;, \quad \ov{b}_{N} \,  =  \, \tau_N b_{N} \quad \e{and} \quad \ov{a}_{N} \,  = \, \tau_N a_{N} \;.
\enq
Moreover, without further notice we shall assume in this subsection that the lower bound $b_N-a_N\, \geq \, \eta$ with $\eta>0$ holds and that $N$ is large enough.

The closed subspace $\mf{X}_s(\R)$ along with the inverse $\mc{W}_N$ are both described in terms of a piecewise holomorphic $2\times 2$ matrix valued function $\chi$
that we shall discuss below. First, however, we provide the expression for the inverse $\mc{W}_N$. For any sufficiently regular $g \in \mf{X}_s(\R)$, the latter takes the form
of the integral transform
\beq
 \mc{W}_N\big[ g \big](\xi) \; = \; \f{ \tau_N^2 }{ 2\pi } \Int{ \R + 2\i\eps^{\prime}}{} \f{ \dd \la }{ 2\i\pi } \Int{ \R + \i \eps^{\prime} }{} \f{\dd \mu }{2\i\pi}
\f{ \ex{-\i \tau_N \la ( \xi - a_{N} ) }  }{ \mu - \la } \Big\{ \chi_{11}(\la) \chi_{12}(\mu) \, - \, \f{\mu}{\la} \chi_{11}(\mu) \chi_{12}(\la) \Big\}
\ex{-\i \ov{b}_{N} \mu} \mc{F}\big[ g \big]( \tau_N \mu)
 \; .
\label{ecriture transformation integrale WN}
\enq
in which $\eps^{\prime}>0$ can be taken as small as need be.

$ \mc{W}_N$ and $\mc{J}$ are both expressed in terms of the unique solution $\chi$ to the $2\times2$ Riemann--Hilbert problem
\begin{itemize}
 \item $\chi_{1a} \in \mc{O}\big(\Cx \setminus \R\big)$ and  $\la \mapsto \la \chi_{2a}(\la) \in \mc{O}\big(\Cx \setminus \R\big)$ admit continuous $\pm$ boundary values on $\R$;
\item $\chi_+(\la) \; = \; G_{\chi}(\la) \, \chi_-(\la)$, with a jump matrix
\beq
G_{\chi}(\la) \; = \; \left( \ba{cc} \ex{\i \la \ov{x}_{N}} & 0   \\
                                        R(\la) \; & \;  -\ex{ - \i \la \ov{x}_{N}} \ea \right) \qquad \e{where} \qquad
R(\la) \; = \; \f{ \sinh\Big[ \tfrac{\la}{2} \Big(\tfrac{1}{\om_1}+\tfrac{1}{\om_2} \Big) \Big]  }
{2  \sinh\Big[ \tfrac{\la}{2\om_1} \Big] \sinh\Big[ \tfrac{\la}{2\om_2} \Big] }
\enq
\item as $\la \tend \infty$
\beq
\chi(\la) \; = \; \left\{ \ba{ccc}
 \left(\ba{cc} -\e{sgn}\Big[ \Re(\la) \Big]\ex{\i \la \ov{x}_{N} }    & 1 \\ -1 & 0  \ea \right) \cdot  \big[ -\i \la \big]^{-\f{\sg_3}{2} }
 \cdot \bigg( I_2 \, + \, \f{ \chi_1}{ \la } \, + \,  \e{O}\Big( \f{1}{\la^2 }\Big) \bigg) \;,  & \la \in \mathbb{H}^+ \vspace{2mm}  \\
 \left(\ba{cc} -1    & \e{sgn}\Big[ \Re(\la) \Big]\ex{-\i \la \ov{x}_{N} } \\ 0 & 1  \ea \right) \cdot  \big[ \i \la \big]^{-\f{\sg_3}{2} } \ex{ \i \f{\pi}{2} \sg_3 }
 \cdot \bigg( I_2 \, + \, \f{ \chi_1}{ \la } \, + \,  \e{O}\Big( \f{1}{\la^2 }\Big) \bigg) \; ,  & \la \in \mathbb{H}^-
\ea \right. \; .
\enq

\end{itemize}

It was established in  \cite{KozBorotGuionnetLargeNBehMulIntOfToyModelSoVType} that $\chi$ satisfies the variable reflection relation
\beq
\chi(-\la)  \; = \; \left( \ba{cc} \chi_{11}(\la)  & - \la \, \chi_{11}(\la) +  \chi_{12}(\la)   \\
                        -\chi_{21}(\la)  &  \la \, \chi_{21}(\la) -  \chi_{22}(\la) \ea \right) \;,
\label{ecriture ppte inversion pour chi}
\enq
the complex conjugation property
\beq
\Big( \chi(\la^*) \Big)^*  \; = \; \left( \ba{cc} - \chi_{11}(-\la)  &   \chi_{12}(-\la)   \\
                        \chi_{21}(-\la)  &   -  \chi_{22}(-\la) \ea \right) \;.
\label{ecriture ppte conjugaison complexe pour chi}
\enq

\begin{figure}[h]
\begin{center}

\includegraphics[width=.7\textwidth]{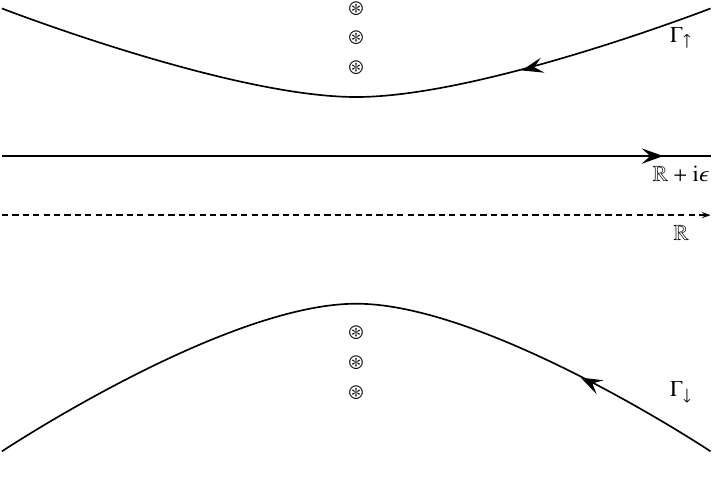}

\caption{Contour $\Ga_{\ua/\da}$ delimiting regions of uniform asymptotic expansion of $\chi$.\label{contours Gamma Up Down pour le RHP de chi}}
\end{center}
\end{figure}

Again, it follows from \cite{KozBorotGuionnetLargeNBehMulIntOfToyModelSoVType} that $\chi$ admits an explicit large-$N$ asymptotic behaviour
valid as soon as $b_N-a_N$ is bounded away from zero uniformly in $N$. Below, we list the uniform  large-$N$ asymptotic expansions in the regions of $\Cx$ which are pertinent for our needs.
These regions are delimited by the real axis and the curves $\Ga_{\ua/\da}$ as depicted in Fig.~\ref{contours Gamma Up Down pour le RHP de chi}.
Note that the point $\i$, resp. $-\i$, may be below or above of $\Ga_{\ua}$, resp. $\Ga_{\da}$, depending on the chosen range for $\om_1, \om_2$.
First, however, we point out that $R$ admits a Wiener-Hopf-like factorisation
\beq
R(\la) \; = \; R_{\ua}(\la) \,  R_{\da}(\la)
\enq
where
\beq
R_{\ua}(\la) =   \f{ \i }{\la} \cdot \sqrt{\om_1+\om_2} \cdot
\bigg( \f{ \om_2 }{  \om_1+\om_2 } \bigg)^{ \f{\i \la }{2 \pi \om_1}  }
\hspace{-3mm} \cdot \;  \bigg( \f{ \om_1 }{  \om_1+\om_2 } \bigg)^{ \f{\i \la }{2 \pi \om_2} }
\hspace{-3mm}  \cdot\;
\f{ \pl{p=1}{2}\Ga\bigg(1-  \f{ \i\la}{2\pi\om_p} \bigg) }{ \Ga\bigg( 1- \f{ \i\la(\om_1 + \om_2) }{2\pi\om_1\om_2} \bigg) }
  \label{ecriture explicite R plus} \\
\enq
and
\beq
R_{\da}(\la) =   \f{ \la }{2\pi \sqrt{\om_1+\om_2} }  \cdot
 \bigg( \f{ \om_2 }{ \om_1+\om_2 } \bigg)^{ -  \f{\i \la }{2 \pi \om_1} }
\hspace{-3mm} \cdot \; \bigg( \f{ \om_1 }{  \om_1+\om_2 } \bigg)^{ -\f{\i \la }{2 \pi \om_2} }
\cdot\f{ \pl{p=1}{2}\Ga\bigg( \f{\i\la}{ 2\pi\om_p } \bigg) }
{ \Ga\bigg( \f{ \i\la(\om_1 + \om_2) }{ 2\pi\om_1\om_2 } \bigg) } \;.
\label{ecriture explicite R moins}
\enq
Note that
\beq
R_{\da}(0) \; = \; - \i  \sqrt{\om_1+\om_2}  \qquad \e{and} \qquad
\Big( \la R_{\ua}(\la) \Big)_{\mid \la=0} \; = \; \i  \sqrt{\om_1+\om_2} \;.
\enq
Also, $R_{\ua}$ and $R_{\da}$ satisfy to the relations
\beq
R_{\ua}(-\la) \; = \; \la^{-1} \cdot R_{\da}(\la) \qquad \e{and}   \qquad  \Big( R_{\ua}(\la^*) \Big)^{*} \; = \; \la^{-1} \cdot R_{\da}(\la) \;.
\label{ecriture identite conjugaison R plus et R moins}
\enq
Furthermore, $R_{\ua/\da}$ admit the asymptotic behaviour
\beqa
R_{\ua}(\la) &  =  & \big( - \i  \la \big)^{-\f{1}{2}} \cdot  \Big( 1\; + \; \e{O}\big( \la^{-1} \big) \Big) \qquad \e{for} \quad
\la \underset{  \la \in  \mathbb{H}^+   }{ \longrightarrow }  \infty   \\
 R_{\da}(\la) & = &   - \i   \big( \i \la \big)^{\f{1}{2}}  \cdot \Big( 1\; + \; \e{O}\big( \la^{-1} \big) \Big) \qquad \e{for} \quad
\la \underset{  \la \in  \mathbb{H}^-   }{ \longrightarrow }  \infty \;.
\eeqa
 The notation $\ua$ and $\da$ indicates the direction, in respect to $\R+\i\eps$ in the complex plane where  $R_{\ua/\da}$ have no pole nor zeroes.

The mentioned uniform asymptotic expansions involve an auxiliary, piecewise analytic, matrix
\beq
\Pi(\la) \, =\, \op{I}_2 \, + \, \e{O}\Big( \f{  \ex{-  \zeta (1-\eta)  \ov{x}_N }   }{ 1 + |\la|} \Big)  \qquad \e{with} \qquad  \zeta \, = \, \f{2\pi \om_1 \om_2}{ \om_1 + \om_2} \;.
\label{ecriture DA de Pi et definition du control LN}
\enq
This bound holds for any fixed $\eta>0$ and uniformly on $\Cx$. Moreover, the remainder is smooth in $a_N$ and $b_N$ with derivatives controlled as
\beq
\Dp{a_N}^k \Dp{b_N}^{\ell}\Pi(\la) \; = \;  \op{I}_2 \de_{k;0}\de_{\ell;0} \, + \, \e{O}\Big( \tau_N^{k+\ell} \f{  \ex{-  \zeta (1-\eta)  \ov{x}_N }   }{ 1 + |\la|} \Big)
\enq

Finally, one has
\beq
\op{P}_{R}(\la) \;  = \; \op{I}_2 \, + \, \f{ \wt{\vth}_{R}  }{ \la } \Pi^{-1}(0) \sg^- \Pi(0) \qquad \e{with} \qquad
 \wt{\vth}_{R}  \; = \; \f{  1 }{ 1 \, - \,  \big[\Pi^{\prime}(0) \Pi^{-1}(0) \big]_{12} } \;.
\label{ecriture operateur PR et def vth R}
\enq

The entries $\chi_{11}, \chi_{12}$ admit holomorphic continuations from $\mathbb{H}^{\mp}$ into some small tubular neighbourhood of $\R$ in $\mathbb{H}^{\pm}$.
In particular, $\chi_{11;\pm}(\la), \chi_{12;\pm}(\la)$ are regular at $\la=0$. In their turn,  $\chi_{11}, \chi_{12}$ admit meromorphic continuations from $\mathbb{H}^{\mp}$
into some small tubular neighbourhood of $\R$ in $\mathbb{H}^{\pm}$. They admit only one pole, which is simple at $\la=0$, and one has the behaviour
\beq
\chi_{21;-}(\la)\; = \;  R_{\da}(0) \wt{\vth}_{R} \Pi_{11}(0) \cdot \f{ 1 }{ \la } \, + \, \e{O}(1) \qquad
\e{and} \qquad
\chi_{22;-}(\la)\; = \;  R_{\da}(0) \wt{\vth}_{R} \Pi_{12}(0) \cdot \f{ 1 }{ \la } \, + \, \e{O}(1)
\enq
as $\la \tend 0^+$.

\subsubsection*{$\bullet$ $\la$ between $\R+\i\eps$ and $\Ga_{\ua}$}

Here, $\eps>0$ is some fixed, small enough, constant. It can be taken so that $\eps^{\prime} > \eps $, with $\eps^{\prime}$ as appearing in \eqref{ecriture transformation integrale WN}.

In this region, it holds that
\beq
\chi(\la) \; = \; \chi_{\infty}(\la) \; + \; \de \chi(\la)
\label{ecriture forme DA chi entre R plus i eps et Gamma up}
\enq
where
\beq
 \chi_{\infty}(\la) \; = \; \left( \ba{cc} \tf{1}{  [\la R_{\ua}(\la) ] } \, - \, \tf{ \ex{\i \la \ov{x}_{N;\a} }}{ R_{\da}(\la) }   & \tf{1}{ R_{\ua}(\la)  } \\
                                    - R_{\ua}(\la)                             &                             0   \ea \right) \;,
\label{ecriture DA dominant chi dans region entre R + i eps et Gamma up}
\enq
while
\bem
\de \chi(\la) \; = \; \left( \ba{c} \Big\{ \tfrac{1}{  \la R_{\ua}(\la)  } \, - \, \tfrac{ \ex{\i \la \ov{x}_{N;\a} }}{ R_{\da}(\la) } \Big\}  \cdot \big[ \de(\Pi \op{P}_{R})(\la) \big]_{11}
\; + \,
\tfrac{ [ \de(\Pi \op{P}_{R})(\la)  ]_{21} }{ R_{\ua}(\la)  }  \vspace{2mm}  \\- R_{\ua}(\la)  \big[ \de(\Pi \op{P}_{R})(\la) \big]_{11} \ea \right. \\
\left. \ba{c} \Big\{ \tfrac{1}{  \la R_{\ua}(\la)  } \, - \, \tfrac{ \ex{\i \la \ov{x}_{N;\a} }}{ R_{\da}(\la) } \Big\}  \cdot \big[ \de(\Pi \op{P}_{R})(\la) \big]_{12}
\; + \,
\tfrac{ [ \de(\Pi \op{P}_{R})(\la)  ]_{22} }{ R_{\ua}(\la)  }  \vspace{2mm} \\
                                    - R_{\ua}(\la)      \big[ \de(\Pi \op{P}_{R})(\la) \big]_{12}                          \ea \right) \;.
\end{multline}
The formulae for the remainder matrix involve
\beq
\de(\Pi \op{P}_{R})(\la) \; = \; \left( \ba{cc} 1 & 0 \\ -\tf{1}{\la} & 1 \ea \right) \cdot \Bigg[ \Pi \op{P}_{R}(\la) \, - \, \left( \ba{cc} 1 & 0 \\ \tf{1}{\la} & 1 \ea \right) \Bigg] \;.
\enq
Finally, one has the uniform entrywise estimate on the remainder
\beq
\de \chi(\la) \; = \; \e{O}\Big( \ex{-  \zeta (1-\eta)  \ov{x}_N } \Big) \;.
\enq

\subsubsection*{$\bullet$ $\la$ between $\Ga_{\da}$ and $\R$}

In this region, it holds that
\beq
\chi(\la) \; = \; \chi_{\infty}(\la) \; + \; \de \chi(\la)
\label{ecriture decomposition chi fct dominante et perturbative sour R}
\enq
where
\beq
 \chi_{\infty}(\la) \; = \; \left( \ba{cc} \tfrac{-1}{   R_{\da}(\la) } \Big( 1 \, - \,  \tfrac{  R_{\da}(\la) }{ \la R_{\ua}(\la) }\ex{-\i \la \ov{x}_{N;\a} } \Big)
 &  \tfrac{ \ex{-\i \la \ov{x}_{N;\a} } }{ R_{\ua}(\la)  } \vspace{2mm} \\
                                     \tfrac{R_{\da}(\la) }{ \la }                             &                            R_{\da}(\la)   \ea \right) \;,
\label{ecriture DA dominant chi dans region sous R}
\enq
while
\bem
\de \chi(\la) \; = \; \left( \ba{c} \tfrac{-1}{   R_{\da}(\la) } \Big( 1 \, - \,  \tfrac{  R_{\da}(\la) }{ \la R_{\ua}(\la) }\ex{-\i \la \ov{x}_{N;\a} } \Big)  \cdot \big[ \de(\Pi \op{P}_{R})(\la) \big]_{11}
\; + \,
\tfrac{ \ex{-\i \la \ov{x}_{N;\a} } }{ R_{\ua}(\la) } \cdot  [ \de(\Pi \op{P}_{R})(\la)  ]_{21}  \vspace{2mm}  \\
\tfrac{R_{\da}(\la) }{ \la }  \cdot    \big[ \de(\Pi \op{P}_{R})(\la) \big]_{11}     +   R_{\da}(\la) \big[ \de(\Pi \op{P}_{R})(\la) \big]_{21}   \ea \right. \\
\left. \ba{c} \tfrac{-1}{   R_{\da}(\la) } \Big( 1 \, - \,  \tfrac{  R_{\da}(\la) }{ \la R_{\ua}(\la) }\ex{-\i \la \ov{x}_{N;\a} } \Big)  \cdot \big[ \de(\Pi \op{P}_{R})(\la) \big]_{12}
\; + \,
\tfrac{ \ex{-\i \la \ov{x}_{N;\a} } }{ R_{\ua}(\la) } \cdot  [ \de(\Pi \op{P}_{R})(\la)  ]_{22}  \vspace{2mm}  \\
\tfrac{R_{\da}(\la) }{ \la }  \cdot    \big[ \de(\Pi \op{P}_{R})(\la) \big]_{12}     +   R_{\da}(\la) \big[ \de(\Pi \op{P}_{R})(\la) \big]_{22}                        \ea \right) \;.
\label{ecriture DA Sous dominant chi dans region sous R}
\end{multline}
One has the uniform entrywise  estimate on the remainder
\beq
\de \chi(\la) \; = \; \e{O}\Big(  \ex{-  \zeta (1-\eta)  \ov{x}_N }  \Big) \;.
\enq
Moreover, one infers that
\beq
\ba{ccc} \chi_{21}(\la) & = & \f{ R_{\da}(0) \wt{\vth}_R \Pi_{11}(0)  }{ \la } \; + \; \e{O}(1)  \vspace{2mm}  \\
\chi_{22}(\la) & = & \f{ R_{\da}(0) \wt{\vth}_R \Pi_{12}(0)  }{ \la } \; + \; \e{O}(1)     \ea   \qquad \e{as} \qquad \la  \tend 0  \qquad \e{with} \qquad \e{Im}(\la)<0 \;.
\label{ecriture residu chi21 et chi22 en zero du plan inf}
\enq
and that $\chi_{11}(\la)$ and $\chi_{12}(\la)$ admit bounded limits as $\la \tend 0$.

\vspace{3mm}

A remark is in order: while both the subspace constraint functional $\mc{J}$ and the inverse $ \mc{W}_N$ involve the Fourier transform of $g$, their
values  \textit{only} depend on the values of $g$ on $\intff{a_N}{b_N}$. This property basically follows from the jump conditions of $\chi$.
It will allow us, later on, to simplify some of the handlings.

\begin{lemme}
\label{Lemma independence valeurs g hoes aN bN}

Let $g_1, g_2 \in H_s(\R)$, $0<s<\tf{1}{2}$ be such that $g_1=g_2$ on $\intff{a_N}{b_N}$. Then,
\beq
 \mc{J} [ g_1  ] \; = \;  \mc{J} [ g_2  ] \qquad and \qquad  \mc{W}_N [ g_1  ] \; = \;  \mc{W}_N [ g_2  ] \;.
\enq

\end{lemme}

\Proof

 Since $g \bs{1}_{\intff{a_N}{b_N}} \in H_s(\R)$, $0<s<\tf{1}{2}$ as soon as $g\in H_s(\R)$, it is enough to show that $\mc{J}[g]$ and $\mc{W}_N[g]$
only depend on $g \bs{1}_{\intff{a_N}{b_N}}$.
We discuss the proof only in the case of sufficiently regular $g$ having fast decay at infinity although this can be done in full generality within the distributional setting of $H_{s}(\R)$
functions.

One may decompose $\mc{J}$ in the form
\beq
 \mc{J} [ g ] \; =   \mc{J}_{L}[ g ] \, + \,  \mc{J}_{\e{c}}[ g ] \, + \,  \mc{J}_{R}[ g ]
\enq
with
\beq
\mc{J}_{L}[ g ] \, =  \hspace{-1mm} \Int{\R+\i\eps^{\prime} }{} \hspace{-1mm} \f{\dd \mu }{2\i\pi} \, \chi_{11}(\mu)
\Int{ -\infty }{ a_{N} } \dd \eta g(\eta) \ex{\i \tau_N \mu(\eta-b_{N}) } \;, \quad
\mc{J}_{R}[ g ] \, =  \hspace{-1mm} \Int{\R+\i\eps^{\prime} }{} \hspace{-1mm} \f{\dd \mu }{2\i\pi} \, \chi_{11}(\mu)
\Int{b_N}{ +\infty } \dd \eta g(\eta) \ex{\i \tau_N \mu(\eta-b_{N}) }
\enq
and $\mc{J}_{\e{c}}[ g ] \, = \, \mc{J}_{\e{c}}\big[ g \bs{1}_{\intff{a_N}{b_N}} \big]$.

We now show the vanishing of $\mc{J}_{L}[ g ]$ and $\mc{J}_{R}[ g ]$. Indeed, by using that $\chi_{11}$ is analytic in the upper half plane,  that
\beq
\chi_{11}(   \mu) \, \leq  \, \f{C }{ |  \mu|^{\tf{1}{2}} } \qquad \e{and} \quad
 \Int{b_N}{ +\infty } \dd \eta g(\eta) \ex{\i \tau_N \mu(\eta-b_{N}) } \, = \,
 \f{ -g(b_N) }{ \i \mu \tau_N } \, - \, \f{1}{  \i \mu \tau_N } \Int{b_N}{ +\infty } \dd \eta g^{\prime}(\eta) \ex{\i \tau_N \mu(\eta-b_{N}) } \; ,
\enq
one observes that one has the uniform bound  on $\mathbb{H}^{+}$
\beq
\bigg|    \chi_{11}(\mu) \Int{b_N}{ +\infty } \dd \eta g(\eta) \ex{\i \tau_N \mu(\eta-b_{N}) }  \bigg| \, \leq \, \f{ C }{  |\mu|^{\tf{3}{2}}  }
\enq
and that the bounded functions appearing above are analytic in $\mathbb{H}^+$. This allows one to deform the integrations from $\R+\i \eps $ to $\R+\i M$ with $M>0$ as large as desired by virtue of Morera's theorem.
Then, one has that
\beq
\big| \mc{J}_{R}\big[ g \big] \big|   =  \bigg| \hspace{-1mm} \Int{\R+\i M}{} \hspace{-1mm} \f{\dd \mu }{2\i\pi} \, \chi_{11}(\mu)
\Int{ b_{N} }{ +\infty } \dd \eta g(\eta) \ex{\i \tau_N \mu(\eta-b_{N}) }    \bigg|    \, \leq \,
C \Int{\R }{}  \f{ \dd s  }{ \big[ s^2 + M^2 \big]^{\f{3}{2}} } \; = \; \f{C}{ \sqrt{M} } \Int{\R }{}  \f{ \dd s  }{ \big[ s^2  +  1 \big]^{\f{3}{2}} } \, \limit{M}{+\infty} \, 0 \;.
\enq

One carries out a similar reasoning regarding to $\mc{J}_{L}[ g ]$. The $\mu^{-1}$ decay rate of the Fourier transform  at infinity
and the existence of continuous $+$ boundary values of $\chi_{11}$ on $\R$ ensure that
\bem
\mc{J}_{L}[ g ] \, =  \hspace{-1mm} \Int{\R}{} \hspace{-1mm} \f{\dd \mu }{2\i\pi} \, \chi_{11;+}(\mu)
\Int{-\infty}{a_N} \dd \eta g(\eta) \ex{\i \tau_N \mu(\eta-b_{N}) } \; = \hspace{-1mm}
\Int{\R}{} \hspace{-1mm} \f{\dd \mu }{2\i\pi} \, \chi_{11;-}(\mu)
\Int{-\infty}{a_N} \dd \eta g(\eta) \ex{-\i \tau_N \mu(a_N-\eta) } \\
\, = \, \Int{\R-\i M}{} \hspace{-1mm} \f{\dd \mu }{2\i\pi} \, \chi_{11}(\mu)
\Int{-\infty}{a_N} \dd \eta g(\eta) \ex{-\i \tau_N \mu(a_N-\eta) }  \; \limit{M}{+\infty} \; 0 \;.
\end{multline}
This entails the claim relative to $\mc{J}[g]$. The results relative to $\mc{W}_N[g]$ are obtained in a very similar fashion. \qed

\subsection{A convenient representation for the inverse acting on $V^{\prime}_{N;\a}$}
\label{SousSection Rep Int inverse sur V prime}

In the general case, one may not expect to be able to simplify $\mc{W}_N[ g ]$ beyond its two-dimensional integral representation. However,
since the potential of interest to us takes a very simple and specific form, such simplifications are possible in the case of $\mc{W}_N[ V_{N;\a}^{\prime} ]$.

We first observe that the expression \eqref{ecriture forme DA chi entre R plus i eps et Gamma up} for  $\chi$ on the line $\R+\i\eps^{\prime}$, $\eps^{\prime}> \eps$ leads to the representation
\beq
\chi_{11}(\la) \; = \; \f{ Q_{\ua}(\la) }{ \la R_{\ua}(\la) } \, - \, \ex{ \i \la \ov{x}_{N} } \f{ Q_{\da}(\la) }{   R_{\da}(\la) }
\qquad \e{with} \qquad
\left\{ \ba{ccc}  Q_{\ua}(\la) & = & 1 + \de \big( \Pi \op{P}_{R} \big)_{11}(\la) \, + \, \la  \de \big( \Pi \op{P}_{R} \big)_{21}(\la) \vspace{2mm} \\
 Q_{\da}(\la) & = & 1 + \de \big( \Pi \op{P}_{R} \big)_{11}(\la)

\ea \right.  \;.
\label{ecriture decomposition efficace of chi11 dans R+i eps}
\enq
There, the functions $Q_{\ua/\da}$ are holomorphic and bounded in the region enclosed by the curves $\Ga_{\ua}$ and $\Ga_{\da}$. Similarly,
\beq
\chi_{12}(\la) \; = \; \f{ \wt{Q}_{\ua}(\la) }{ \la R_{\ua}(\la) } \, - \, \ex{ \i \la \ov{x}_{N} } \f{ \wt{Q}_{\da}(\la) }{   R_{\da}(\la) }
\qquad \e{with} \qquad
\left\{ \ba{ccc}  \wt{Q}_{\ua}(\la) & = & \la + \de \big( \Pi \op{P}_{R} \big)_{12}(\la) \, + \, \la  \de \big( \Pi \op{P}_{R} \big)_{22}(\la) \vspace{2mm} \\
 \wt{Q}_{\da}(\la) & = &   \de \big( \Pi \op{P}_{R} \big)_{12}(\la)

\ea \right. \;.
\label{ecriture decomposition efficace of chi12 dans R+i eps}
\enq

For further purpose, it is convenient to introduce the vectors
\beq
\bs{E}_L(\la) \, = \, \left( \ba{cc}  \chi_{11}(\la) \\ - \f{  \chi_{12}(\la) }{ \la } \ea \right)
\;, \quad
\bs{E}_R(\la)\, = \, \left( \ba{cc}  \chi_{12}(\la) \\  \la\,\chi_{11}(\la) \ea \right)
\enq
so that
\beq
\Big( \bs{E}_L(\la), \bs{E}_R(\mu) \Big) \; = \;  \chi_{11}(\la) \chi_{12}(\mu) \, - \, \f{\mu}{\la} \chi_{11}(\mu) \chi_{12}(\la) \;.
\enq
The decompositions \eqref{ecriture decomposition efficace of chi11 dans R+i eps}-\eqref{ecriture decomposition efficace of chi12 dans R+i eps}
for $\chi_{11}$ and $\chi_{12}$ entail that $\bs{E}_R(\la)\, = \, \bs{E}_R^{(\ua)}(\la)\, - \, \ex{\i \la \ov{x}_N }  \bs{E}_R^{(\da)}(\la)$ with
\beq
 \bs{E}_R^{(\ua)}(\la) \, = \, \f{ 1 }{ \la R_{\ua}(\la) }\left( \ba{cc} \wt{Q}_{\ua}(\la) \vspace{1mm} \\ \la Q_{\ua}(\la) \ea \right)  \;, \quad
  \bs{E}_R^{(\da)}(\la) \, = \, \f{ 1 }{   R_{\da}(\la) }\left( \ba{cc} \wt{Q}_{\da}(\la) \vspace{1mm} \\ \la Q_{\da}(\la) \ea \right) \;,
\enq
and similarly $\bs{E}_L(\la)\, = \, \tfrac{1}{\la} \cdot \bs{E}_L^{(\ua)}(\la)\, - \,   \tfrac{ \ex{\i \la \ov{x}_N } }{\la} \cdot \bs{E}_L^{(\da)}(\la)$
\beq
 \bs{E}_L^{(\ua)}(\la) \, = \, \f{ 1 }{ \la R_{\ua}(\la) }\left( \ba{cc} \la Q_{\ua}(\la) \vspace{1mm} \\ - \wt{Q}_{\ua}(\la) \ea \right)  \;, \quad
  \bs{E}_L^{(\da)}(\la) \, = \, \f{ 1 }{   R_{\da}(\la) }\left( \ba{cc} \la Q_{\da}(\la)  \vspace{1mm} \\ -\wt{Q}_{\da}(\la) \ea \right) \; .
\enq
Finally, we also set
\beqa
 \msc{U}_{12}(\la) & = & \f{ N u_{N} \la + \i v_{N} }{ 1+\la^2 } \i \chi_{11}(\i) \;,  \label{definition U12}\\
\msc{U}_{11}(\la) & = & \f{- \chi_{12}(\i)  }{ 1+\la^2 } \cdot \Big( \i  N u_{N} +  \la v_{N}  \Big)
\, - \, \f{ \i \chi_{11}(\i)  }{ \i + \la }   \f{    N u_{N} -  v_{N}  }{ 2} \;,
\label{definition U11}
\eeqa
in which we have introduced the shorthand notations
\beq
 N u_{N} \, = \, \ex{\ov{b}_{N}} \,+ \,\ex{-\ov{a}_{N}}  \qquad \e{and} \qquad v_{N} \; = \;  \ex{\ov{b}_{N}} \,- \,\ex{-\ov{a}_{N}}  \;.
\label{ecriture lien entre bN aN et uN et vN}
\enq

\begin{prop}
\label{ecriture expression exacte pour inverse sur VN alpha prime}

One has the decomposition $\mc{W}_N[ V_{N;\a}^{\prime} ] \, = \, \sul{a=1}{3}\varpi_{N }^{(a)} $ where
\beq
 \varpi_{N }^{  (1)}(\xi) \, = \, \f{ \mf{r} \tau_N }{ 4\i\pi N  } \hspace{-2mm} \Int{ \R + 2\i\eps^{\prime} }{} \hspace{-2mm} \f{ \dd \la }{ 2\i\pi }
 \ex{-\i \tau_N \la (\xi-a_{N}) }
\Big\{ \f{ \chi_{12}(\la) }{ \la } \msc{U}_{12}(\la) \; + \; \chi_{11}(\la) \msc{U}_{11}(\la) \Big\}   \;,
\label{ecriture rep int varpi 1}
\enq
and
\beq
 \varpi_{N }^{   (2)}(\xi) \, = \, \f{ \a \tau_N }{ 2 \i \pi N  } \hspace{-2mm} \Int{ \R + 2\i\eps^{\prime} }{} \hspace{-2mm} \f{ \dd \la }{ 2\i\pi }
 \ex{-\i \tau_N \la (\xi-a_{N}) }  \f{ \chi_{11}(\la) \chi_{12;-}(0)  }{\la}  \;.
\label{ecriture rep int varpi 2}
\enq

Finally, it holds that  $\varpi_{N }^{(3)}(\xi)\,= \,  \varpi_{N;\ua }^{   (3)}(\xi) \, + \, \varpi_{N;\da }^{   (3)}(\xi) \, + \, \varpi_{N;0 }^{   (3)}(\xi)$.
The building blocks of this decomposition take the form, for $\ups \in \{\ua, \da\}$,
\beqa
 \varpi_{N; \ups }^{   (3)}(\xi)   =    \f{  \eps_{\ups} \tau_N }{ 4\i\pi N  } \hspace{-2mm} \Int{ \R + 2\i\eps^{\prime} }{} \hspace{-2mm} \f{ \dd \la }{ 2\i\pi } \ex{-\i \tau_N \la (\xi-a_{N}) }
\Big( \bs{E}_L(\la), \bs{\mc{E}}^{(\ups)}(\la) \Big)    \qquad with   \qquad \eps_{\ups}=\left\{ \ba{cc} 1 & \ups = \ua \\ -1 & \ups = \da  \ea \right.  \;.
\eeqa
Here, we have set
\beqa
 \bs{\mc{E}}^{(\ua)}(\la) & =  &  \ex{- \ov{b}_N   } \f{2 \mc{F}[\op{g}](\i) }{ \pi (\la+\i)}  \bs{E}_R^{(\ua)}(-\i) \bs{1}_{1<\zeta}\, +
  \Int{ \R -  \i \varkappa_{\eta}  }{} \hspace{-2mm} \f{ \dd \mu }{ 2\i\pi }
 \f{ \mu \mc{F}[\op{g}](\mu)  \, \ex{-\i \ov{b}_N \mu  } }{ (  \mu - \la )  \cosh\Big[ \tfrac{\pi \mu }{2} \Big] }  \bs{E}_R^{(\ua)}(\mu)  \\
 \bs{\mc{E}}^{(\da)}(\la) & =  &  \ex{  \ov{a}_N   } \f{2 \mc{F}[\op{g}](\i) }{ \pi (\i-\la)}  \bs{E}_R^{(\da)}(\i) \bs{1}_{1<\zeta}\, +
  \Int{ \R +  \i \varkappa_{\eta}  }{} \hspace{-2mm} \f{ \dd \mu }{ 2\i\pi }
 \f{ \mu \mc{F}[\op{g}](\mu)  \, \ex{-\i \ov{a}_N \mu  } }{ (  \mu - \la )  \cosh\Big[ \tfrac{\pi \mu }{2} \Big] }  \bs{E}_R^{(\da)}(\mu)
\eeqa
while
\beq
 \varpi_{N; 0 }^{   (3)}(\xi) \, =  \,  \f{  -\tau_N }{ 4\i\pi N  } \hspace{-2mm} \Int{ \R + 2\i\eps^{\prime} }{} \hspace{-2mm} \f{ \dd \la }{ 2\i\pi } \ex{-\i \tau_N \la  \xi  }
\Big( \bs{E}_L(\la), \bs{E}^{(\da)}_R(\la) \Big)  \f{ \la \mc{F}[\op{g}](\la)  }{  \cosh\Big[ \tfrac{\pi \la }{2} \Big] } \;.
\enq
Above, we have introduced
\beq
\varkappa_{\eta} \, = \, (1-\eta) \min \big\{ 2, \zeta \big\} \qquad  with  \qquad \zeta \, = \, 2\pi \f{\om_1 \om_2 }{ \om_1 + \om_2 } \;.
\label{definition varkappa eta et zeta}
\enq

\end{prop}

\Proof

Owing to Lemma \ref{Lemma independence valeurs g hoes aN bN}, one may choose $V^{\prime}_{N;\a}$ to take any values outside of $\intff{a_N}{b_N}$ so as to compute the
Fourier transform occurring in the expression for $\mc{W}_N$, provided the function belongs to $H_s(\R)$ with $0<s<\tf{1}{2}$. Thus, we choose to extend $V^{\prime}_{N;\a}$ from $\intff{a_N}{b_N}$ to $\R$ as
$V_{N;\a}(\la) \, = \, \mf{v}_{N;\a}(\la)  \bs{1}_{ \intff{a_N}{b_N} }(\la) \, + \,  \mf{w}_{N}(\la) $, where
\beq
\mf{v}_{N;\a}(\la) \, = \, \f{ \mf{r} }{ N \tau_N} \cosh\big[ \tau_N \la \big]  \, - \, \f{ \a \la   }{ N }  \qquad \e{and} \qquad
 \mf{w}_{N}(\la)   \, = \, \Int{\R}{} \f{ \dd \mu }{2\pi N } \cdot \f{ \op{g}( \tau_N \mu)  }{  \cosh\big[ \tau_N (\la-\mu) \big] } \;.
\enq
A direct calculation yields
\beq
\Int{ a_{N} }{ b_{N} } \!   \dd s \, \ex{\i\mu \tau_N (s - b_{N}) } \mf{v}_{N;\a}^{\prime}(s)  \; = \;
\f{ \mf{r} }{ 2 \tau_N N \i} \sul{ \sg = \pm }{} \sg \f{ \ex{\sg \ov{b}_{N} } \, - \, \ex{-\i \mu \ov{x}_{N} } \ex{\sg \ov{a}_{N} } }{ \mu - \i\sg }
\; - \;  \a \f{ 1 \, - \, \ex{ -\i \mu \ov{x}_{N} } }{ \i\mu N \tau_N  } \;.
\enq
Further, one has that
\bem
\Int{ \R }{}\dd s  \ex{\i \mu \tau_N s } \mf{w}_{N}^{\prime}(s) \; = \; -\i\mu \tau_N \Int{\R}{} \dd s \mf{w}_{N} (s)\ex{\i \mu \tau_N s }  \\
\; = \; -\f{ \i\mu \tau_N}{2 \pi N } \Int{\R}{} \dd  s \f{ \ex{\i \mu \tau_N s } }{ \cosh(\tau_N s) } \cdot \Int{\R}{} \dd t  \ex{\i \mu \tau_N t } \op{g}(\tau_N t)
\; = \; -  \f{\i \mu \, \mc{F}[\op{g}](\mu) }{ 2 \tau_N N \cosh\Big[ \f{\pi \mu }{2} \Big] } \;.
\label{calcul TF facteur wn prime}
\end{multline}
As a result, one obtains a decomposition  $\mc{W}_N[ V_{N;\a}^{\prime} ] \, = \, \sul{a=1}{3}\varpi_{N;\a}^{(a)} $ where
\beqa
\varpi_{N }^{(1)}(\xi)  & = &   \f{ \tau_N \mf{r} }{ 4\i \pi N } \hspace{-2mm} \Int{ \R + 2\i\eps^{\prime} }{} \hspace{-2mm} \f{ \dd \la }{ 2\i\pi } \Int{ \R + \i \eps^{\prime} }{} \hspace{-2mm} \f{\dd \mu }{2\i\pi}
\f{ \ex{-\i \tau_N \la ( \xi - a_{N} ) }  }{ \mu - \la } \Big\{ \chi_{11}(\la) \chi_{12}(\mu) \, - \, \f{\mu}{\la} \chi_{11}(\mu) \chi_{12}(\la) \Big\}
  \sul{ \sg = \pm }{} \sg \f{ \ex{\sg \ov{b}_{N} } \, - \, \ex{-\i \mu \ov{x}_{N} } \ex{\sg \ov{a}_{N} } }{ \mu - \i\sg } \;, \nonumber \\
\varpi_{N }^{(2)}(\xi) & = &  - \f{ \tau_N \a }{ 2\i \pi N } \hspace{-2mm} \Int{ \R + 2\i\eps^{\prime} }{} \hspace{-2mm} \f{ \dd \la }{ 2\i\pi } \Int{ \R + \i \eps^{\prime} }{} \hspace{-2mm} \f{\dd \mu }{2\i\pi}
\f{ \ex{-\i \tau_N \la ( \xi - a_{N} ) }  }{ \mu - \la } \Big\{ \chi_{11}(\la) \chi_{12}(\mu) \, - \, \f{\mu}{\la} \chi_{11}(\mu) \chi_{12}(\la) \Big\}
  \f{ 1 \, - \, \ex{ -\i \mu \ov{x}_{N} } }{  \mu    }\;, \nonumber
\eeqa
and
\beq
\varpi_{N }^{(3)}(\xi) \; = \; -\f{ \i \tau_N  }{ 4\pi N } \hspace{-2mm} \Int{ \R + 2\i\eps^{\prime} }{} \hspace{-2mm} \f{ \dd \la }{ 2\i\pi } \hspace{-2mm}
\Int{ \R + \i \eps^{\prime} }{} \hspace{-2mm} \f{\dd \mu }{2\i\pi}
\f{ \ex{-\i \tau_N \la ( \xi - a_{N} ) }  }{ \mu - \la } \Big\{ \chi_{11}(\la) \chi_{12}(\mu) \, - \, \f{\mu}{\la} \chi_{11}(\mu) \chi_{12}(\la) \Big\}
  \ex{ - \i  \mu \ov{b}_{N} } \f{  \mu \, \mc{F}[\op{g}](\mu) }{   \cosh\Big[ \f{\pi \mu }{2} \Big] }  \; .
\enq

We first compute the $\mu$-integral arising in $\varpi_{N }^{(1)}$ and $\varpi_{N }^{(2)}$.  For such a purpose, one starts by observing that $\chi_{1a}$ admits an analytic continuation from $\mathbb{H}^{-}$
to $\mathbb{H}^+$. Denoting this analytic continuation as $\chi_{1a;-}$, one has the relation
\beq
\chi_{1a;-}(\la) \; = \; \chi_{1a}(\la) \ex{-\i \la \ov{x}_{N} } \quad \e{with} \quad \Im[\la] >0 \;.
\enq
  Then, one may express $\varpi_{N }^{(1)}$ in the form
\bem
\varpi_{N }^{(1)}(\xi)  \,  = \,    \f{ \tau_N \mf{r} }{ 4\i \pi N } \hspace{-2mm} \Int{ \R + 2\i\eps^{\prime} }{} \hspace{-2mm} \f{ \dd \la }{ 2\i\pi } \hspace{-2mm}
              \Int{ \R + \i \eps^{\prime} }{} \hspace{-2mm} \f{\dd \mu }{2\i\pi}
\f{ \ex{-\i \tau_N \la ( \xi - a_{N} ) }  }{ \mu - \la } \Big\{ \chi_{11}(\la) \chi_{12}(\mu) \, - \, \f{\mu}{\la} \chi_{11}(\mu) \chi_{12}(\la) \Big\}
  \sul{ \sg = \pm }{}  \f{ \sg \ex{\sg \ov{b}_{N} }   }{ \mu - \i\sg } \\
 - \f{ \tau_N \mf{r} }{ 4\i \pi N } \hspace{-2mm} \Int{ \R + 2\i\eps^{\prime} }{} \hspace{-2mm} \f{ \dd \la }{ 2\i\pi } \hspace{-2mm} \Int{ \R + \i \eps^{\prime} }{} \hspace{-2mm} \f{\dd \mu }{2\i\pi}
\f{ \ex{-\i \tau_N \la ( \xi - a_{N} ) }  }{ \mu - \la } \Big\{ \chi_{11}(\la) \chi_{12;-}(\mu) \, - \, \f{\mu}{\la} \chi_{11;-}(\mu) \chi_{12}(\la) \Big\}
  \sul{ \sg = \pm }{} \f{ \sg   \ex{\sg \ov{a}_{N} } }{ \mu - \i\sg }
\end{multline}
We could split the integral in two pieces since each integrand behaves at infinity as $\e{O}\big(|\mu|^{-\tf{3}{2}}\big)$. Then, because of these bounds,
one may take the first $\mu$-integral by means of the residues of the poles located above the line $\R+\i \eps^{\prime}$ and
take the second $\mu$-integral by means of the residues of the poles located below the line $\R+\i \eps^{\prime}$. Note that there is no pole at
$\mu-\la$ in the first integral so that only the pole at $\mu=\i$ contributes, while, in the second case, only the pole at $\mu=-\i$ does.
This yields
\beq
 \varpi_{N }^{  (1)}(\xi) \, = \, \f{ \mf{r} \tau_N }{ 4\i\pi N  } \hspace{-2mm} \Int{ \R + 2\i\eps^{\prime} }{} \hspace{-2mm} \f{ \dd \la }{ 2\i\pi }
 \ex{-\i \tau_N \la (\xi-a_{N}) }
 \sul{\sg = \pm}{} \ex{ \ov{b}_{N}^{(\sg)} }
 \f{  \chi_{11}(\la) \chi_{12}(\sg \i ) \, - \, \tf{\sg \i \chi_{11}(\sg \i) \chi_{12}(\la) }{ \la }  }{ \i - \sg \la }
\enq
in which we have used the shorthand notations
\beq
\ov{b}_{N}^{(+)}\; = \; \ov{b}_{N} \qquad   \e{and}   \qquad
\ov{b}_{N}^{(-)}\; = \; - \ov{a}_{N} \;.
\enq
Then, it is a matter of direct calculation to observe that owing to the inversion relation \eqref{ecriture ppte inversion pour chi}
one gets
\beq
\sul{\sg = \pm}{}  \ex{ \ov{b}_{N;\a}^{(\sg)} }  \f{  \chi_{11}(\la) \chi_{12}(\sg \i )
                    \, - \, \tf{\sg \i \chi_{11}(\sg \i) \chi_{12}(\la) }{ \la }  }{ \i - \sg \la } \,  = \, \f{ \chi_{12}(\la) }{ \la } \msc{U}_{12}(\la) \; + \; \chi_{11}(\la) \msc{U}_{11}(\la)
\label{decomposition X via U11 et U12}
\enq
with $\msc{U}_{1a}$ as given in \eqref{definition U12}-\eqref{definition U11}. This was the last step before reaching \eqref{ecriture rep int varpi 1}.

\vspace{1mm}

The same reasoning yields \eqref{ecriture rep int varpi 2} starting from the previous expression for $\varpi_{N  }^{(2)}$:
the part of the integral deformed up to $+\i \infty$ produces $0$
while the part deformed to $-\i\infty$ picks a simple pole at $\mu=0$.

\vspace{1mm}

We now turn on to rewriting $\varpi_{N  }^{(3)}$, which can be recast as
\beq
\varpi_{N }^{(3)}(\xi)  \,  =     \f{ \tau_N  }{ 4\i \pi N } \hspace{-2mm}\Int{ \R + 2\i\eps^{\prime} }{}  \hspace{-2mm} \f{ \dd \la }{ 2\i\pi } \Int{ \R + \i \eps^{\prime} }{} \hspace{-2mm}  \f{\dd \mu }{2\i\pi}
\f{ \ex{-\i \tau_N \la ( \xi - a_{N} ) }  }{ \mu - \la } \bigg\{ \ex{-\i \mu \ov{b}_N} \Big( \bs{E}_L(\la), \bs{E}_R^{(\ua)}(\mu) \Big)
\, - \, \ex{-\i \mu \ov{a}_N} \Big( \bs{E}_L(\la), \bs{E}_R^{(\da)}(\mu) \Big)   \bigg\}
\f{  \mu \, \mc{F}[\op{g}](\mu) }{  \cosh\Big[ \f{\pi \mu }{2} \Big] } \;.
\enq
One then splits the integral in two pieces, one containing $\bs{E}_R^{(\ua)}$ and the other one $\bs{E}_R^{(\da)}$. Both integrals are well defined due to
the $\e{O}(\mu^{-k})$ behaviour at infinity of $ \mc{F}[\op{g}](\mu)$ ensured by the $\e{O}\big( \ex{- \f{\mf{r}}{2} \cosh(\xi) } \big)$ control on $\op{g}^{(k)}(\xi)$
for any $k\geq 0$ and the $\e{O}\big( |\la|^{-\tf{3}{2}} \big)$ behaviour of the integrand pointwise in $\mu$.
Then, in the integral involving $\bs{E}_R^{(\ua)}$, one moves the $\mu$ integration from $\R + \i \eps^{\prime}$ to
$\R - \i \varkappa_{\eta} $. There are four potential sources of poles in the integrand
\beq
\f{1}{R_{\ua}(\mu)} \;\;   \e{poles}\, \e{at} \;\; \mu = - \i \zeta n \;, \quad
\f{1}{R_{\da}(\mu)} \; \;   \e{poles}\, \e{at} \; \;  \mu =   \i \zeta n \;, \quad
 \f{1}{ \cosh\big[ \tfrac{\pi \mu}{2} \big] }\; \;    \e{poles}\, \e{at} \; \; \mu = \pm \i (1+2n)
\enq
 with $n \in \mathbb{N}^{*}$, and $\mu = \la$.  Thus, in deforming the integration contour, provided that $1<\zeta$, one picks up a pole at $\mu= - \i$, else no poles are crossed.
 This then yields $\varpi_{N;\ua}^{(3)}(\xi)$

Similarly, in the second integral involving $\bs{E}_R^{(\da)}$, one moves the $\mu$-integration contour from $\R + \i \eps^{\prime}$
to $\R + \i \varkappa_{\eta} $. This produces one contribution stemming from the pole at $\mu=\la$ and one contribution
stemming from the pole at $\mu=\i$. The last contribution is only present if $1<\zeta$. The terms obtained in this way correspond to
$\varpi_{N, \da }^{(3)}(\xi) $ for the $\mu$-integrals over $\R +\i \varkappa_{\eta} $ and $\varpi_{N, 0 }^{(3)}(\xi) $ for the residue at $\mu=\la$ part. \qed

\subsection{Support of the equilibrium measure}
\label{SousSection support mesure eq}

When constructing the equilibrium measure, on top of determining its density, one also needs to fix its support.
Since the density belongs to $H_{s}(\R)$, see \cite{KozBorotGuionnetLargeNBehMulIntOfToyModelSoVType}, with $0<s < \tf{1}{2}$,
and satisfies the singular integral equation \eqref{ecriture eqn int sing pour densite} throughout its support $\intff{ a_{N;\a} }{ b_{N;\a} }$
which satisfies owing to Lemma \ref{Lemme borne inf sup sur tailler du support de la mesure equilibre} the lower-bound $ b_{N;\a} \, - \, a_{N;\a} \geq 2\vsg>0$,
one gets that, for any $N$ large enough, one has the representation
\beq
\wh{\varrho}_{\e{eq};\a} \; =\;\mc{W}_N\big[ V_{N;\a}^{\prime} \big]_{\mid a_N, b_N \hookrightarrow a_{N;\a}, b_{N;\a} }  \;.
\label{ecriture expression explicite pour densite mesure equilibire}
\enq
We stress that the inverse operator $\mc{W}_N$ given in \eqref{ecriture transformation integrale WN} is now subordinate to the yet unknown pair of points $ a_{N;\a},\, b_{N;\a}$ delimiting the support.
One then gets two additional constraints, the first one translating the fact that   $V_{N;\a}^{\prime} \in \mc{S}_N\big[ H_{s}(\intff{ a_{N;\a} }{ b_{N;\a} }) \big]$ with $0<s<\tf{1}{2}$,
and the second one expressing the unit mass property of the measure
\beq
0 \; = \; \mc{J}\big[ V_{N;\a}^{\prime}  \big]_{\mid   a_N,   b_N    \hookrightarrow    a_{N;\a},   b_{N;\a}    }  \qquad \e{and} \qquad
1 \; = \;  \Int{ a_{N;\a} }{ b_{N;\a} } \dd \xi \Big\{ \mc{W}_N\big[ V_{N;\a}^{\prime} \big] (\xi)  \Big\}_{\mid   a_N,   b_N    \hookrightarrow    a_{N;\a},   b_{N;\a}    } \;.
\label{ecriture contraintes sur pts bord aN et bN}
\enq
In this subsection, we shall establish that the constraints \eqref{ecriture contraintes sur pts bord aN et bN}
admit a unique solution, for $N$ large enough, provided that $ a_{N;\a} \, \leq \, - \vsg$ and $ b_{N;\a} \, \geq \,   \vsg$,
a property that is ensured by Lemma \ref{Lemme borne inf sup sur tailler du support de la mesure equilibre}.
Hence, this solution does provide one with the support of the equilibrium measure.
We close the subsection by establishing the explicit form of the first few terms in the large-$N$ expansion of $a_{N;\a}$ and $b_{N;\a}$.
This ends the proof of Theorem \ref{Theoreme principal}.

\begin{prop}
\label{Proposition contrainte cpte en aN et bN explicitee en DA}

For given endpoints  $a_N, b_N$ satisfying $x_N>\eta$ for some $\eta>0$, the constraint functional $\mc{I}_1\big[ V^{\prime}_{N;\a}\big]$ defined in \eqref{ecriture fnelle contraintes}
admits the large-$N$ asymptotic expansion
\beq
\mc{J}\big[ V^{\prime}_{N;\a}\big] \; = \; \f{ \mf{r}  \chi_{11}(i) }{ 2\i N \tau_N} \big( \ex{\ov{b}_{N} } \, -\,\ex{-\ov{a}_{N} }    \big)
\; - \;  \f{ \a  \chi_{11;-}(0) }{  \i N \tau_N}
- \, \f{  \mc{F}[\op{g}](\i)  }{ \pi N \tau_N R_{\ua}(-\i) } \Big\{   \ex{- \ov{b}_N} \, - \,   \ex{ \ov{a}_N }  \Big\}\bs{1}_{1<\zeta}   \; + \; \mf{R}_{\mc{J}}(a_N, b_N)
\enq
with a remainder that is smooth in $a_N, b_N$ and controlled as
\beq
\Dp{a_N}^k \Dp{b_N}^{\ell} \mf{R}_{\mc{J}}(a_N, b_N)  \; = \; \e{O}\bigg( \tau^{k+\ell}_N  \f{ \ex{- \ov{b}_N \varkappa_{\eta}  } + \ex{  \ov{a}_N \varkappa_{\eta}  } }{ N \tau_N  } \bigg)  \;.
\enq
  $\varkappa_{\eta}$ appearing above is as introduced in \eqref{definition varkappa eta et zeta} \;.

\end{prop}

\Proof

One consecutively computes each of the contributions to $\mc{J}$. The one of $\mf{v}_{N;\a}$ can be obtained in closed form. Indeed, one has
\bem
\mc{J}\big[ \mf{v}_{N;\a}^{\prime} \big] \, =   \Int{\R + \i\eps^{\prime} }{} \hspace{-2mm} \f{ \dd \mu }{2\i\pi N \tau_N } \chi_{11}(\mu)
\bigg\{ \sul{ \sg = \pm }{} \f{  \mf{r} \sg \ex{\sg \ov{b}_{N} }   }{ 2\i (\mu - \i\sg) }
\; - \;   \f{ \a  }{ \i\mu   } \bigg\}  \\
\, -   \Int{\R + \i\eps^{\prime} }{}  \hspace{-2mm} \f{ \dd \mu }{2\i\pi \tau_N N } \chi_{11}(\mu) \ex{-\i \mu \ov{x}_{N} }
\bigg\{  \sul{ \sg = \pm }{}   \f{   \mf{r} \sg  \ex{\sg \ov{a}_{N} } }{ 2  \i (\mu - \i\sg)  }
\; - \;  \f{   \a   }{ \i\mu  } \bigg\} \;.
\end{multline}
Note that, each integrand is a $\e{O}\big( \mu^{-\tf{3}{2}} \big)$ at $\infty$. The first integral can be computed by taking the residues of the poles located above of
$\R + \i\eps^{\prime}$. There is a simple pole at $\mu = \i$. To compute the second integral, one observes that $\chi_{11}$ admits an analytic continuation from $\mathbb{H}^{-}$
to $\mathbb{H}^+$. Denoting this analytic continuation as $\chi_{11;-}$, it holds  $\chi_{11;-}(\la) \; = \; \chi_{11}(\la) \ex{-\i \la \ov{x}_{N} } $
with $\Im[\la] >0$. Thus, in the second integral, one replaces $\chi_{11}$ with $\chi_{11;-}$ and then takes the integral in terms of the residues at the poles located below of
$\R + \i\eps^{\prime}$. There are two poles, one simple at $\mu=\i$ and one simple at $\mu=0$. This yields
\beqa
\mc{J}\big[ \mf{v}_{N;\a}^{\prime} \big] & =  &  \f{ \mf{r} }{ 2\i \tau_N N } \Big( \chi_{11}(\i)  \ex{  \ov{b}_N } \, - \,  \chi_{11}(-\i)  \ex{ - \ov{a}_N } \Big)
\, - \, \f{   \a  \chi_{11;-}(0)   }{ \i\tau_N N }  \\
& = & \f{ \mf{r}\chi_{11}(\i)   }{ 2\i \tau_N N } \Big(  \ex{ \ov{b}_N } \, - \,  \ex{ - \ov{a}_N } \Big)
\, - \, \f{   \a  \chi_{11;-}(0)   }{ \i\tau_N N }   \;.
\eeqa
Here, we have simplified the expression owing to \eqref{ecriture ppte conjugaison complexe pour chi}.

Further, observe that integrations by parts and the $\e{O}\big( \ex{- \mf{r} \cosh(\xi) } \big)$ decay of $\op{g}^{(k)}(\xi)$ for any $k \geq 0 $ ensure that $\mc{F}[\op{g}](\mu)= \e{O}\big( \mu^{-k} \big)$. Then,
recalling the representations \eqref{ecriture decomposition efficace of chi11 dans R+i eps}, \eqref{calcul TF facteur wn prime} and inserting  into the expression for the constraint functional leads to
\bem
\mc{J} \big[ \mf{w}_{N}^{\prime} \big] \,  = \,    \Int{\R + \i\eps^{\prime} }{} \hspace{-2mm} \f{ \dd \mu }{4\i^2\pi N \tau_N }
  \f{ Q_{\ua}(\mu) \mc{F}[\op{g}](\mu) }{  R_{\ua}(\mu)  \cosh\big[ \tfrac{\pi \mu }{2} \big] } \ex{-\i \mu\ov{b}_{N} }
 \,  - \,   \Int{\R + \i\eps^{\prime} }{} \hspace{-2mm} \f{ \dd \mu }{4\i^2\pi N \tau_N }
\f{ \mu Q_{\da}(\mu) \mc{F}[\op{g}](\mu) }{  R_{\da}(\mu)  \cosh\big[ \tfrac{\pi \mu }{2} \big] } \ex{-\i \mu \ov{a}_{N} }  \\
\, = \, \Int{\R - \i \varkappa_{\eta}  }{} \hspace{-2mm} \f{ \dd \mu }{4\i^2\pi N \tau_N }
  \f{ Q_{\ua}(\mu) \mc{F}[\op{g}](\mu) }{  R_{\ua}(\mu)  \cosh\big[ \tfrac{\pi \mu }{2} \big] } \ex{-\i \mu\ov{b}_{N} }
 \,  - \,   \Int{\R + \i \varkappa_{\eta} }{} \hspace{-2mm} \f{ \dd \mu }{4\i^2\pi N \tau_N }
\f{ \mu Q_{\da}(\mu) \mc{F}[\op{g}](\mu) }{  R_{\da}(\mu)  \cosh\big[ \tfrac{\pi \mu }{2} \big] } \ex{-\i \mu \ov{a}_{N} } \\
\, - \, \f{  \mc{F}[\op{g}](\i)  }{ \pi N \tau_N R_{\ua}(-\i) } \Big\{ Q_{\ua}(-\i) \ex{- \ov{b}_N} \, - \,   Q_{\da}(\i) \ex{ \ov{a}_N }  \Big\}\bs{1}_{1<\zeta} \;.
\end{multline}
In the second line we have deformed the integration contour to $\R - \i \varkappa_{\eta}$ for the first integral and to $\R + \i \varkappa_{\eta}$.
Note that, in the process, one only picks poles of $\tf{1}{ \cosh\big[ \tfrac{\pi \mu}{2} \big] }$ at $\pm \i $, and this provided that $1<\zeta$. This generates the contribution of the last line,
which also uses the inversion relations for $R_{\ua/\da}$ and the parity of $\mc{F}[\op{g}]$.

Owing to the $\e{O}(\mu^{-k})$ for any $k\geq 0$ decay of the integrand at infinity, one readily then estimates the first integral to be $\e{O}\Big( \tf{ \ex{- \varkappa_{\eta} \ov{b}_N } }{ N  \tau_N } \Big)$
and the second one to be $\e{O}\Big( \tf{ \ex{  \varkappa_{\eta} \ov{a}_N } }{ N  \tau_N } \Big)$. Finally, one has $Q_{\ua/\da}(\mu)\, = \, 1\, + \, \e{O}\big( \ex{-\zeta(1-\eta) \ov{x}_N} \big)$
uniformly in $\mu \in  \R \pm \i \varkappa_{\eta}$ and for $\mu=\pm \i$. Thus, all-in-all,
\beq
\mc{J} \big[ \mf{w}_{N}^{\prime} \big] \,  = \,   - \, \f{  \mc{F}[\op{g}](\i)  }{ \pi N \tau_N R_{\ua}(-\i) } \Big\{   \ex{- \ov{b}_N} \, - \,   \ex{ \ov{a}_N }  \Big\}\bs{1}_{1<\zeta}
\; +  \f{1}{N \tau_N} \e{O}\Big(  \big[ \ex{- \ov{b}_N} \, + \,   \ex{ \ov{a}_N } \big] \ex{-\zeta(1-\eta) \ov{x}_N} \, + \, \ex{- \varkappa_{\eta} \ov{b}_N } \, + \, \ex{  \varkappa_{\eta} \ov{a}_N } \Big) \;.
\enq
It is clear from the previous handlings and the properties of $Q_{\ua/\da}$ that the remainder is smooth in $a_N, b_N$ and that each $a_N$ or $b_N$ derivative
of the remainder worsens the control by a factor of $\tau_N$.

The result then follows upon putting together all the estimates and exact expressions.  \qed





\begin{prop}
\label{Proposition calcul points de bord aN et bN}

It holds
\bem
\Int{ a_{N} }{ b_{N} } \dd \xi \,\mc{W}_N\big[ V_{N;\a}^{\prime} \big] (\xi)\;= \; \f{- \a }{2\pi N}  \chi_{11;-}^{\prime}(0)\chi_{12;-}(0)
\, -\, \f{ \mf{r}   }{ 4\i \pi N } \bigg\{ \i \Big( \ex{\ov{b}_{N} } \,- \,\ex{-\ov{a}_{N} } \Big) \Big(   \f{ \chi_{11;-}(0) \chi_{11}(\i)  }{ 2}  - \chi_{12;-}^{\prime}(0) \chi_{11}(\i) \Big)  \\
\, + \, \big( \ex{ \ov{b}_{N} } \,+ \,\ex{ - \ov{a}_{N} }    \big) \big[ \chi_{11;-}(0) \chi_{12}(\i) \,- \, \chi_{12;-}(0) \chi_{11}(\i) \, - \,
\f{\i}{2}  \chi_{11;-}(0) \chi_{11}(\i) \big] \bigg\} \\
\; + \; \f{ \mc{F}[\op{g}](\i) }{ 2\pi^2 N R_{\da}(\i) R_{\da}(0) }  \big( \ex{-\ov{b}_N}\,+ \, \ex{\ov{a}_N} \big) \bs{1}_{1<\zeta}
\; + \;
\e{O}\bigg( \f{ \ex{\varkappa_{\eta}\ov{a}_N} \, +\, \ex{-\varkappa_{\eta}\ov{b}_N}  }{N}      \bigg) \;.
\end{multline}
  $\varkappa_{\eta}$ appearing above is as introduced in \eqref{definition varkappa eta et zeta} \;.

\end{prop}

\Proof

One starts from the partially integrated expression for $\mc{W}_N\big[ V_{N;\a}^{\prime} \big]$ obtained in Proposition \ref{ecriture expression exacte pour inverse sur VN alpha prime}.
Then, with the notation of that proposition, one has
\beq
\Int{ a_{N} }{ b_{N} } \dd \xi \,\mc{W}_N\big[ V_{N;\a}^{\prime} \big] (\xi)\;= \; \sul{a=1}{3} \Int{ a_{N} }{ b_{N} } \dd \xi \,\varpi_N^{(a)} (\xi) \;.
\enq
Since,
\beq
\Int{a_N}{b_N} \dd \xi \ex{-\i \la \tau_N (\xi-a_N)} \; = \; \f{1 \, - \, \ex{-\i \la \ov{x}_N } }{ \i \la \tau_N }\;,
\label{ecriture expression integral de Fourier de exp sur support mesure eq}
\enq
one gets that
\beq
 \Int{ a_{N} }{ b_{N} } \dd \xi \,\varpi_N^{(2)} (\xi)  \, = \,-  \f{ \a   \chi_{12;-}(0) }{ 2  \pi N  } \hspace{-2mm} \Int{ \R + 2\i\eps^{\prime} }{} \hspace{-2mm} \f{ \dd \la }{ 2\i\pi }
  \f{ \chi_{11}(\la)\, - \, \chi_{11;-}(\la)   }{\la^2 } \;.
\enq
There, we have used that $\chi_{1a}$ admits and analytic continuation from $\mathbb{H}^{-}$ to $\mathbb{H}^{+}$ denoted $\chi_{1a;-}(\la) \; = \;  \ex{-\i \la \ov{x}_N } \chi_{1a}(\la) $
for $\Im(\la)>0$. The integral is well defined in that the integrand behaves as $\e{O}(|\la|^{-\tf{5}{2}})$ at $\infty$. Splitting it in two pieces and
taking the integral involving $\chi_{11}$ by means of the residues at the poles located above $\R + 2\i\eps^{\prime}$ -there are none-
and the integral involving $\chi_{11;-}$ by means of the residues at the poles located below of $\R + 2\i\eps^{\prime}$ -there is only one at $\la=0$-
one gets that
\beq
 \Int{ a_{N} }{ b_{N} } \dd \xi \,\varpi_N^{(2)} (\xi)  \, = \,-  \f{ \a   \chi_{12;-}(0) }{ 2  \pi N  } \Dp{\la}\chi_{11;-}(0) \;.
\enq

We next focus on the contribution involving  $\varpi_{N }^{  (1)}$. One gets
\beq
 \Int{ a_{N} }{ b_{N} } \dd \xi \,\varpi_N^{(1)} (\xi)  \, = \, - \f{ \mf{r}   }{ 4 \pi N  } \hspace{-2mm} \Int{ \R + 2\i\eps^{\prime} }{} \hspace{-2mm} \f{ \dd \la }{ 2\i\pi }
\bigg\{ \f{ \chi_{12}(\la) -  \chi_{12;-}(\la)}{ \la^2 } \msc{U}_{12}(\la) \; + \; \f{ \chi_{11}(\la) -\chi_{11;-}(\la) }{ \la } \msc{U}_{11}(\la) \bigg\}   \;.
\enq
Splitting it in two pieces and
taking the integral involving $\chi_{11}$ by means of the residues at the poles located above $\R + 2\i\eps^{\prime}$ -there are none-
and the integral involving $\chi_{11;-}$ by means of the residues at the poles located below of $\R + 2\i\eps^{\prime}$ -there is only one at $\la=0$-
one gets that
\beq
 \Int{ a_{N} }{ b_{N} } \dd \xi \,\varpi_N^{(1)} (\xi)  \, = \,- \f{ \mf{r}   }{ 4 \pi N  } \Big\{ \Dp{\la} \big( \chi_{12;-} \msc{U}_{12} \big)(0)
 \, + \, \chi_{11;-}(0)   \msc{U}_{11}(0) \Big\} \;.
\enq
A direct calculation  leads to
\beq
  \msc{U}_{11}(0) \, = \, - \i N u_N \chi_{12}(\i) \, - \, \chi_{11}(\i) \f{ N u_N - v_N}{2} \;, \quad
\msc{U}_{12}(0) \, = \,  -   v_N \chi_{11}(\i)  \quad \e{and} \quad
\msc{U}_{12}^{\prime}(0) \, = \,  \i N u_N \chi_{11}(\i)  \; .
\enq
Here, we remind that $u_N$ and $v_N$ have been introduced in \eqref{ecriture lien entre bN aN et uN et vN}.
Upon inserting the above  into the closed expression for the integral of $\varpi_N^{(1)}$, one eventually gets
\bem
 \Int{ a_{N} }{ b_{N} } \dd \xi \,\varpi_N^{(1)} (\xi)  \, = \,\f{ \mf{r}   }{ 4 \pi N  } \bigg\{  \i N u_N \Big[   \chi_{11;-}(0) \chi_{12}(\i)  \, - \,  \chi_{12;-}(0) \chi_{11}(\i)
 \, - \, \f{\i}{2} \chi_{11;-}(0) \chi_{11}(\i)  \Big] \\
\, - \, v_N \Big[ \f{1}{2} \chi_{11;-}(0) \chi_{11}(\i) \, - \,   \chi_{12;-}^{\prime}(0) \chi_{11}(\i) \Big]  \bigg\} \;.
\end{multline}

Finally, we focus on estimating the contribution issuing from $\varpi_N^{(3)}$. Starting from the expression for the integrand provided in Proposition \ref{ecriture expression exacte pour inverse sur VN alpha prime}
along with \eqref{ecriture expression integral de Fourier de exp sur support mesure eq}, one gets that
\beq
 \Int{ a_{N} }{ b_{N} } \dd \xi \,\varpi_N^{(3)} (\xi)  \, = \, \mc{I}^{(3)}_{\da} \, + \, \mc{I}^{(3)}_{\ua} \, + \, \mc{I}^{(3)}_{0} \;.
\enq
There, one has
\bem
\mc{I}^{(3)}_{\da}   \, =  \,  \f{  1 }{ 4 \pi N  } \hspace{-2mm} \Int{ \R + 2\i\eps^{\prime} }{} \hspace{-2mm} \f{ \dd \la }{ 2\i\pi \la^2 }
\Big(   \big[ \bs{E}_L^{(\ua)}(\la) \, + \, \big(1  - \ex{\i \la \ov{x}_N }  \big) \bs{E}_L^{(\da)}(\la)     \big] \, - \,   \ex{ -\i \la \ov{x}_N }\bs{E}_L^{(\ua)}(\la)  , \bs{\mc{E}}^{(\da)}(\la) \Big) \\
\, =  \,  \f{  1 }{ 4 \pi N  } \hspace{-2mm} \Int{ \R + \i\varkappa_{\eta^{\prime}} }{} \hspace{-2mm} \f{ \dd \la }{ 2\i\pi \la^2 }
\Big(   \bs{E}_L^{(\ua)}(\la)   \, + \, \big(1  -   \ex{\i \la \ov{x}_N }  \big) \bs{E}_L^{(\da)}(\la)     , \bs{\mc{E}}^{(\da)}(\la) \Big)
\, + \, \bs{1}_{1<\zeta}  \f{  \ex{\ov{a}_N }  \mc{F}[\op{g}](\i)   }{2 \pi^2 N }  \Big(  \bs{E}_L^{(\ua)}(\i)  , \bs{E}_R^{(\da)}(\i) \Big)   \\
+ \;  \f{  1 }{ 4 \pi N  } \Dp{\la} \bigg\{  \ex{-\i \la \ov{x}_N }  \Big(\bs{E}_L^{(\ua)}(\la)  , \bs{\mc{E}}^{(\da)}(\la) \Big) \bigg\}_{\mid \la=0}
\, - \,  \f{  1 }{ 4 \pi N  } \hspace{-2mm} \Int{ \R - \i\varkappa_{\eta^{\prime}}  }{} \hspace{-2mm} \f{ \dd \la }{ 2\i\pi \la^2 }   \ex{ -\i \la \ov{x}_N }
\Big(  \bs{E}_L^{(\ua)}(\la)  , \bs{\mc{E}}^{(\da)}(\la) \Big) \;.
\end{multline}
Above, $\eta^{\prime}>\eta$, and is taken small enough. Also, we have made use of the relation  $\Big(  \bs{E}_L^{(\da)}(\i)  , \bs{E}_R^{(\da)}(\i) \Big)=0$.

To estimate the various contributions more precisely, one needs the auxiliary estimates
\beq
\Big(  \bs{E}_L^{(\ua)}(\la)  , \bs{E}_R^{(\da)}(\mu) \Big) \; = \; \f{- \mu }{ R_{\ua}(\la) R_{\da}(\mu)  } \; + \; \e{O}\Big(  \sqrt{ (1+|\la|) (1+|\mu|) } \cdot  \ex{- \zeta (1-\eta) \ov{x}_N } \Big)
\enq
and
\beq
\Big(  \bs{E}_L^{(\da)}(\la)  , \bs{E}_R^{(\da)}(\mu) \Big) \; = \;   \e{O}\Big(  \sqrt{ (1+|\la|) (1+|\mu|) } \cdot  \ex{- \zeta (1-\eta) \ov{x}_N } \Big) \;.
\enq
Note that the remainders are holomorphic in $\la$ located between $\Ga_{\ua}$ and $\Ga_{\da}$ and smooth in $a_N, b_N$, with each derivative adding a $\tau_N$ factor to the control.
From that and the fact that $\mc{F}[\op{g}]$ is a Schwartz function, one infers the bounds
\beq
\Big(  \bs{E}_L^{(\ua)}(\la)  , \bs{\mc{E}}^{(\da)}(\la) \Big) \; = \; - \ex{\ov{a}_N} \f{2 \i \mc{F}[\op{g}](\i) }{ \pi (\i-\la) R_{\ua}(\la) R_{\da}(\i) } \bs{1}_{1<\zeta}
\; + \; \f{1}{R_{\ua}(\la)}\e{O}\bigg(  \f{  \ex{  \ov{a}_N \varkappa_{\eta} } }{ \sqrt{ 1+|\la|} }\bigg)
\; + \; \e{O}\bigg(  \f{  \ex{  \ov{a}_N \varkappa_{\eta} }  \ex{-\zeta(1-\eta)\ov{x}_N } }{ \sqrt{ 1+|\la|} }\bigg)
\enq
so that the first remainder has a zero at $\la=0$, and
and
\beq
\Big(  \bs{E}_L^{(\da)}(\la)  , \bs{\mc{E}}^{(\da)}(\la) \Big) \; = \;   \e{O}\bigg(  \f{  \ex{   \ov{a}_N \varkappa_{\eta} } }{ \sqrt{ 1+|\la|} }\bigg) \;.
\enq
There, the remainders enjoy the same properties as above. Inserting these bounds inside of the obtained representation for $\mc{I}_{\da}^{(3)}$, one gets
\beq
\mc{I}_{\da}^{(3)} \, = \,   \ex{\ov{a}_N} \f{ \mc{F}[\op{g}](\i) }{ 2 \pi^2 N R_{\da}(\i) } \bs{1}_{1<\zeta}   \bigg\{ \f{ 1 }{ \i R_{\ua}(\i)  } \, + \, \f{1}{R_{\da}(0) } \bigg\} \; + \;
\e{O}\bigg(  \f{  \ex{  \ov{a}_N \varkappa_{\eta} } }{ N }\bigg) \;.
\enq
The remainder is now only $\mc{C}^{1}$ in respect to $a_N, b_N$ and partial $a_N$ or $b_N$ derivatives thereof enjoy the same control with a $\tau_N$
additional factor.

Similarly, one obtains
\bem
\mc{I}^{(3)}_{\ua} \, = \, \f{ - 1 }{ 4 \pi N  } \hspace{-2mm} \Int{ \R + 2\i\eps^{\prime} }{} \hspace{-2mm} \f{ \dd \la }{ 2\i\pi \la^2 }
\Big( \big[ \bs{E}_L^{(\ua)}(\la) \, + \, \big(1  - \ex{\i \la \ov{x}_N }  \big) \bs{E}_L^{(\da)}(\la)     \big] \, - \,   \ex{ -\i \la \ov{x}_N }\bs{E}_L^{(\ua)}(\la), \bs{\mc{E}}^{(\ua)}(\la) \Big)  \\
\, =  \,  \f{  -1 }{ 4 \pi N  } \hspace{-2mm} \Int{ \R + \i\varkappa_{\eta^{\prime}} }{} \hspace{-2mm} \f{ \dd \la }{ 2\i\pi \la^2 }
\Big(   \bs{E}_L^{(\ua)}(\la)   \, + \, \big(1  -   \ex{\i \la \ov{x}_N }  \big) \bs{E}_L^{(\da)}(\la)     , \bs{\mc{E}}^{(\ua)}(\la) \Big)
\, - \, \bs{1}_{1<\zeta}  \f{  \ex{-\ov{b}_N -\ov{x}_N }  \mc{F}[\op{g}](\i)   }{2 \pi^2 N }  \Big(  \bs{E}_L^{(\ua)}(-\i)  , \bs{E}_R^{(\da)}(-\i) \Big)   \\
- \;  \f{  1 }{ 4 \pi N  } \Dp{\la} \bigg\{  \ex{-\i \la \ov{x}_N }  \Big(\bs{E}_L^{(\ua)}(\la)  , \bs{\mc{E}}^{(\ua)}(\la) \Big) \bigg\}_{\mid \la=0}
\, + \,  \f{  1 }{ 4 \pi N  } \hspace{-2mm} \Int{ \R - \i\varkappa_{\eta^{\prime}}  }{} \hspace{-2mm} \f{ \dd \la }{ 2\i\pi \la^2 }   \ex{ -\i \la \ov{x}_N }
\Big(  \bs{E}_L^{(\ua)}(\la)  , \bs{\mc{E}}^{(\ua)}(\la) \Big) \;.
\end{multline}
As before, one gets the auxiliary estimates
\beq
\Big(  \bs{E}_L^{(\ua)}(\la)  , \bs{E}_R^{(\ua)}(\mu) \Big) \; = \; \e{O}\Big(  \sqrt{ (1+|\la|) (1+|\mu|) } \cdot  \ex{- \zeta (1-\eta) \ov{x}_N } \Big)
\enq
and
\beq
\Big(  \bs{E}_L^{(\da)}(\la)  , \bs{E}_R^{(\ua)}(\mu) \Big) \; = \;  \f{ \la  }{ R_{\da}(\la) R_{\ua}(\mu)  } \; + \;   \e{O}\Big(  \sqrt{ (1+|\la|) (1+|\mu|) } \cdot  \ex{- \zeta (1-\eta) \ov{x}_N } \Big) \;.
\enq
Again, the remainders are holomorphic in $\la$ located between $\Ga_{\ua}$ and $\Ga_{\da}$ and smooth in $a_N, b_N$, with each derivative adding a $\tau_N$ factor to the control.
Thus,
\beq
\Big(  \bs{E}_L^{(\ua)}(\la)  , \bs{\mc{E}}^{(\ua)}(\la) \Big) \; = \;
 \e{O}\bigg(  \f{  \ex{  -\ov{b}_N  -\zeta(1-\eta)\ov{x}_N }   }{ \sqrt{ 1+|\la|} }\bigg)
\enq
and
\beq
\Big(  \bs{E}_L^{(\da)}(\la)  , \bs{\mc{E}}^{(\ua)}(\la) \Big) \; = \;   \f{2  \la \mc{F}[\op{g}](\i) \,  \ex{-\ov{b}_N} }{ \pi (\i+\la) R_{\da}(\la) R_{\ua}(-\i) } \bs{1}_{1<\zeta}
\, + \, \e{O}\bigg(  \f{  \ex{   -\ov{b}_N \varkappa_{\eta} } }{ \sqrt{ 1+|\la|} }\bigg) \;.
\enq
There, the remainders enjoy the same properties as above. Hence, one gets that
\beq
\mc{I}^{(3)}_{\ua}   \, =  \, - \f{ \mc{F}[\op{g}](\i) \,  \ex{-\ov{b}_N} } { 2 \pi^2 N  R_{\ua}(-\i) } \bs{1}_{1<\zeta}
\Int{ \R +  \i\varkappa_{\eta} }{} \hspace{-2mm} \f{ \dd \la }{ 2\i\pi \la  R_{\da}(\la) (\i+\la) }
 \; + \; \e{O}\bigg( \f{  \ex{-\varkappa_{\eta}\ov{b}_N } }{N}  \bigg) \;.
\enq
The remainder is $\mc{C}^{1}$ in respect to $a_N, b_N$ and partial $a_N$ or $b_N$ derivatives thereof enjoy the same control with a $\tau_N$
additional factor. The remaining integral can be computed by means of taking the residues at $\la=0$ and $\la=-\i$ located below of $ \R +  \i\varkappa_{\eta}$, leading
eventually to
\beq
\mc{I}^{(3)}_{\ua}   \, =  \,  \ex{-\ov{b}_N} \f{ \mc{F}[\op{g}](\i) }{ 2 \pi^2 N R_{\da}(\i) } \bs{1}_{1<\zeta}   \bigg\{ \f{ 1 }{ \i R_{\ua}(\i)  } \, + \, \f{1}{R_{\da}(0) } \bigg\}
 \; + \; \e{O}\bigg( \f{  \ex{-\varkappa_{\eta}\ov{b}_N } }{N}  \bigg) \;.
\enq

It remains to focus on $\mc{I}^{(3)}_{0}$ which takes the form
\bem
\mc{I}^{(3)}_{0}   \, =  \,  \f{  1 }{ 4\pi N  } \hspace{-2mm} \Int{ \R + 2\i\eps^{\prime} }{} \hspace{-2mm} \f{ \dd \la }{ 2\i\pi \la }
\Big( \ex{-\i \la \ov{a}_N }   \big[ \bs{E}_L^{(\ua)}(\la) \, + \, \big(1  - \ex{\i \la \ov{x}_N }  \big) \bs{E}_L^{(\da)}(\la)     \big]
\, - \,   \ex{ -\i \la \ov{b}_N }\bs{E}_L^{(\ua)}(\la), \bs{E}^{(\da)}_R(\la) \Big)  \f{  \mc{F}[\op{g}](\la)  }{  \cosh\Big[ \tfrac{\pi \la }{2} \Big] }  \\
\; = \;  \f{   1}{ 4\pi N  } \hspace{-2mm} \Int{ \R +  \i\varkappa_{\eta} }{} \hspace{-2mm} \f{ \dd \la }{ 2\i\pi \la }\ex{-\i \la \ov{a}_N }
\Big(    \bs{E}_L^{(\ua)}(\la)   ,  \bs{E}^{(\da)}_R(\la) \Big)  \f{   \mc{F}[\op{g}](\la)  }{  \cosh\Big[ \tfrac{\pi \la }{2} \Big] }
\, - \, \ex{\ov{a}_N } \f{ \mc{F}[\op{g}](\i)  }{ 2 \pi^2 N }\Big(    \bs{E}_L^{(\ua)}(\i)   ,  \bs{E}^{(\da)}_R(\i) \Big) \bs{1}_{1<\zeta}  \\
 + \, \f{ \mc{F}[\op{g}](0) }{ 4\pi N } \Big(\bs{E}_L^{(\ua)}(0), \bs{E}^{(\da)}_R(0) \Big)
 \, - \, \ex{-\ov{b}_N } \f{ \mc{F}[\op{g}](\i)  }{ 2 \pi^2 N }\Big(    \bs{E}_L^{(\ua)}(-\i)   ,  \bs{E}^{(\da)}_R(-\i) \Big) \bs{1}_{1<\zeta} \\
\, - \, \f{  1 }{ 4\pi N  } \hspace{-2mm} \Int{ \R -   \i\varkappa_{\eta}  }{} \hspace{-2mm} \f{ \dd \la }{ 2\i\pi \la }   \ex{ -\i \la \ov{b}_N }
\Big(\bs{E}_L^{(\ua)}(\la), \bs{E}^{(\da)}_R(\la) \Big)  \f{  \mc{F}[\op{g}](\la)  }{  \cosh\Big[ \tfrac{\pi \la }{2} \Big] } \;.
\end{multline}
It is then enough to invoke the previous auxiliary bounds to infer that $\Big(\bs{E}_L^{(\ua)}(0), \bs{E}^{(\da)}_R(0) \Big) \, = \, \e{O}\big(\mc{L}_N \big)$ while the two integral terms are a
$N^{-1} \e{O}\big( \ex{-\varkappa_{\eta} \ov{b}_N } + \ex{ \varkappa_{\eta} \ov{a}_N } \big)$. Those auxiliary bounds also allow one to simplify the explicit contributions
so that, up to subdominant corrections,
\beq
\mc{I}^{(3)}_{0}   \, =  \,   \f{ \i \mc{F}[\op{g}](\i)  }{ 2 \pi^2 N R_{\ua}(\i) R_{\da}(\i)} \bs{1}_{1<\zeta} \Big( \ex{-\ov{b}_N } + \ex{\ov{a}_N } \Big)
\; + \;   \e{O}\bigg(  \f{ \ex{-\varkappa_{\eta} \ov{b}_N } + \ex{ \varkappa_{\eta} \ov{a}_N } }{ N }\bigg) \;.
\enq
The remainder is $\mc{C}^{1}$ in respect to $a_N, b_N$ and partial $a_N$ or $b_N$ derivatives thereof enjoy the same control with a $\tau_N$
additional factor.  By putting the three estimates together, we get
\beq
 \Int{ a_{N} }{ b_{N} } \dd \xi \,\varpi_N^{(3)} (\xi)  \, = \,  \f{   \mc{F}[\op{g}](\i) \bs{1}_{1<\zeta}  }{ 2 \pi^2 N R_{\da}(0) R_{\da}(\i)} \Big( \ex{-\ov{b}_N } + \ex{\ov{a}_N } \Big)
 \; +\; \e{O}\bigg(  \f{ \ex{-\varkappa_{\eta} \ov{b}_N } + \ex{ \varkappa_{\eta} \ov{a}_N } }{ N }\bigg)\;.
\enq
This entails the claim. \qed

\vspace{2mm}

Below, we establish the unique solvability of the constraints \eqref{ecriture contraintes sur pts bord aN et bN} on the endpoints $a_N, b_N$ under the hypothesis that
$b_N-a_N \geq \eta$, for some fixed $\eta >0$.  By the previous discussion, this implies that these unique solutions do correspond to the endpoints of the support of the equilibrium measure.

\begin{prop}

 Consider the subset of $\R^2$
\beq
\mc{D}_{ \vsg } \, = \,   \intfo{\vsg}{+\infty} \, \times \,  \intof{-\infty}{-\vsg} \;,
\label{definition domaine D vsg}
\enq
with $\vsg >0$ and small. For any $\vsg >0$ there exists $N_0$ such that, for any $N \geq N_0$, there exists a unique solution $\big(b_{N;\a}, a_{N;\a} \big)\in \mc{D}_{\vsg}$
to the constraint equations \eqref{ecriture contraintes sur pts bord aN et bN}. Moreover, it holds that

\beq
b_{N;\a} \, = \, 1 \, + \, \e{o}(1) \qquad and \qquad  a_{N;\a} \, = \, -1 \, + \, \e{o}(1) \;.
\enq

\end{prop}

\Proof

It follows from Propositions \ref{Proposition contrainte cpte en aN et bN explicitee en DA}-\ref{Proposition calcul points de bord aN et bN}
and the expansion $\chi_{11}(\i) \, =\,  \tf{ -\i }{ R_{\ua}(\i) } \, + \, \e{O}\big( \ex{- \ov{x}_N \varkappa_{\eta}} \big)$
with a differentiable remainder, that the constraints \eqref{ecriture contraintes sur pts bord aN et bN} are equivalent to the system of equations for $a_N, b_N$:
\beq
\ex{\ov{b}_{N} } \, -\,\ex{-\ov{a}_{N} }
 \; = \; 2 \f{  \a  \chi_{11;-}(0) }{ \mf{r}   \chi_{11}(i) } \; + \; \e{O}\Big(     \ex{- \ov{b}_N \wt{\varkappa}_{\eta}  } + \ex{  \ov{a}_N \wt{\varkappa}_{\eta}  }   \Big)  \;,
\enq
and
\bem
 \ex{\ov{b}_{N} } \, + \,\ex{-\ov{a}_{N} } \;= \; \f{ 4 \pi N }{ \i \, \mf{r}   } \cdot
\f{  1 \, + \, \f{ \a }{2\pi N}  \Big[ \chi_{11;-}^{\prime}(0)\chi_{12;-}(0)\, - \,  \chi_{12;-}^{\prime}(0)\chi_{11;-}(0) \, + \, \f{1}{2}  \chi_{11;-}^2(0)  \Big] }
{ \chi_{11;-}(0) \chi_{12}(\i) \,- \, \chi_{12;-}(0) \chi_{11}(\i) \, - \, \f{\i}{2}  \chi_{11;-}(0) \chi_{11}(\i)   }  \\
\; + \; \e{O}  \Big( \ex{ \wt{\varkappa}_{\eta}\ov{a}_N } \, +\, \ex{ - \wt{\varkappa}_{\eta}\ov{b}_N }   \Big)   \;.
\end{multline}
There, we agree upon $\wt{\varkappa}_{\eta}=\min \big\{ 1, (1-\eta)\zeta \big\}$.

A direct calculation based on the expansions \eqref{ecriture forme DA chi entre R plus i eps et Gamma up} and \eqref{ecriture decomposition chi fct dominante et perturbative sour R}
leads to
\beq
\chi_{11;-}(0) \chi_{12}(\i) \,- \, \chi_{12;-}(0) \chi_{11}(\i) \, - \, \f{\i}{2}  \chi_{11;-}(0) \chi_{11}(\i) \;= \; \f{-1}{R_{\da}(0) R_{\ua}(\i) }
\bigg( 1 \, + \, \i \ex{-\ov{x}_N}  \f{ R_{\ua}(\i) }{ R_{\da}(\i) } \, +  \, \e{O}\Big( \ex{-\zeta (1-\eta) \ov{x}_N } \Big) \bigg)
\enq
and
\beq
\chi_{11;-}^{\prime}(0)\chi_{12;-}(0)\, - \,  \chi_{12;-}^{\prime}(0)\chi_{11;-}(0) \, + \, \f{1}{2}  \chi_{11;-}^2(0)   \; = \;
\e{O}\Big( \ov{x}_N  \,  \ex{-\zeta (1-\eta) \ov{x}_N } \Big) \;.
\enq
This allows one to recast the second constraint in the form
\beq
 \ex{\ov{b}_{N} } \, + \,\ex{-\ov{a}_{N} } \; = \; N  \mf{c}_0  \f{  1\,  +\, \tfrac{\ov{x}_N}{N}
 \e{O}\Big( \ex{ \wt{\varkappa}_{\eta}\ov{a}_N} \, +\, \ex{- \wt{\varkappa}_{\eta}\ov{b}_N}  \, + \,  \ex{-\zeta (1-\eta) \ov{x}_N }  \Big)  }
 {   1 \, + \, \i \ex{-\ov{x}_N}  \f{ R_{\ua}(\i) }{ R_{\da}(\i) } \, +  \, \e{O}\Big( \ex{-\zeta (1-\eta) \ov{x}_N } \Big)   }
\, = \,  N  \mf{c}_0  \cdot \bigg( 1\,  +\,  \e{O}\Big(   \ex{ \wt{\varkappa}_{\eta}\ov{a}_N} \, +\,  \ex{- \wt{\varkappa}_{\eta}\ov{b}_N}   \Big) \bigg) \;.
\label{ecriture eqn contrainte masse ss forme DA}
\enq
There, we have set
\beq
\mf{c}_0 \, = \, \f{4\pi }{ \mf{r} } \sqrt{\om_1+\om_2} R_{\ua}(\i) \;.
\enq

\noindent Finally, owing to
\beq
 2 \f{  \a  \chi_{11;-}(0) }{ \mf{r}   \chi_{11}(\i) } \; = \; \f{  \a \mf{c}_0 }{ \pi (\om_1+\om_2) }
\f{  1\,  +\,   \e{O}\Big(    \ex{-\zeta (1-\eta) \ov{x}_N }  \Big)  }
 {   1 \, - \, \i \ex{-\ov{x}_N}  \f{ R_{\ua}(\i) }{ R_{\da}(\i) } \, +  \, \e{O}\Big( \ex{-\zeta (1-\eta) \ov{x}_N } \Big)   }
\enq
one recasts the first constrain in the form
\bem
\ex{\ov{b}_{N} } \, -\,\ex{-\ov{a}_{N} }
 \; = \; \f{  \a \mf{c}_0 }{ \pi (\om_1+\om_2) }  \f{  1\,  +\,   \e{O}\Big(    \ex{-\zeta (1-\eta) \ov{x}_N }  \Big)  }
 {   1 \, - \, \i \ex{-\ov{x}_N}  \f{ R_{\ua}(\i) }{ R_{\da}(\i) }    }
 \; + \; \e{O}\bigg(    \ex{- \ov{b}_N \wt{\varkappa}_{\eta}  } + \ex{  \ov{a}_N \wt{\varkappa}_{\eta}  }   \bigg)   \\
\; = \; \f{  \a \mf{c}_0 }{ \pi (\om_1+\om_2) } \,  +\,  \e{O}\Big( \ex{ \wt{\varkappa}_{\eta}\ov{a}_N} \, +\, \ex{- \wt{\varkappa}_{\eta}\ov{b}_N}   \Big) \;.
\label{ecriture eqn contrainte copt local en endpoints ss forme DA}
\end{multline}

In order to prove more efficiently the existence and uniqueness for $N$ large enough of the system's solutions on the domain $\mc{D}_{\vsg}$ introduced in \eqref{definition domaine D vsg},
it is convenient to pass to the finite in $N$-variables $\big( u_N, v_N \big)$ defined through \eqref{ecriture lien entre bN aN et uN et vN}.

Note that upon defining
\beq
\Psi\big( x, y \big) \; = \; \bigg( \f{  \ex{\tau_N x } + \ex{-\tau_N y }  }{ N}   \, , \,  \ex{\tau_N x } - \ex{-\tau_N y }  \bigg) \;,
\enq
one has that
\beq
\wh{\mc{D}}_{\vsg} \; = \; \Psi\big( \mc{D}_{\vsg}  \big) \; = \;  \bigg\{ (u,v) \in \R^+\times \R\; : \;  u \, \geq \, \f{1}{N^{1-\vsg}} \;\; \e{and} \;\;
 N u - 2 N^{\vsg} \geq |v|   \bigg\} \;.
\enq
Then, one may recast the constraints in the form
\beq
u_N \, = \, \mf{c}_0\, + \, \de \Phi_1\big( u_N, v_N\big) \qquad \e{and} \qquad
v_N \, = \, \f{ \a \mf{c}_0 }{ \pi (\om_1+\om_2) }  \, + \, \de \Phi_2\big( u_N, v_N\big) \;.
\enq
The functions $\de \Phi_a$ are $\mc{C}^1$ on $\wh{\mc{D}}_{\vsg} $ since the remainders in \eqref{ecriture eqn contrainte masse ss forme DA} and \eqref{ecriture eqn contrainte copt local en endpoints ss forme DA}
are $\mc{C}^1$ on $\mc{D}_{\vsg}$. Moreover, it is direct to estimate that throughout $\wh{\mc{D}}_{\vsg}$, it holds
\beq
\de \Phi_a\big( u, v\big) \, =  \, \e{O}\bigg(  \Big( \tfrac{2}{N u+v} \Big)^{ \wt{\varkappa}_{\eta } ) } \, + \, \Big( \tfrac{2}{N u -v} \Big)^{ \wt{\varkappa}_{\eta }  } \bigg)
\; = \; \e{O}\Big( N^{ - \vsg\wt{\varkappa}_{\eta }  }  \Big) \;.
\enq
 Taken that the remainder's estimates also hold for the first derivatives up to additional $\tau_N$ factors, one gets that
\beq
\big| \Dp{u}\de \Phi_a\big( u, v\big) \big| \, + \,  \big| \Dp{v}\de \Phi_a\big( u, v\big) \big| \, =  \,\e{O}\Big( N^{ - \vsg\wt{\varkappa}_{\eta }(1-\eta^{\prime}) }  \Big) \;,
\enq
with $\eta^{\prime}>0$ and small enough. Thus, introducing the $\mc{C}^1$ diffeomorphism on $\wh{\mc{D}}_{\vsg}$
\beq
\bs{\Phi}(u,v) \; = \; \left( \ba{cc}  \mf{c}_0 - u  + \de \Phi_1\big( u, v\big) \\
                                \tfrac{ \a \mf{c}_0 }{ \pi (\om_1+\om_2) } - v  + \de \Phi_2\big( u, v\big)  \ea \right)
\enq
one has, uniformly throughout  $\wh{\mc{D}}_{\vsg}$,  that $\op{D}\bs{\Phi} \, = \, - \op{I}_2 \, +\, \e{O}\Big( N^{ - \vsg\wt{\varkappa}_{\eta }(1-\eta^{\prime}) }  \Big)$.
One is thus in the setting where one can invoke the local inversion theorem so as to ensure that $\bs{\Phi}$, as soon as $N$ is large enough,
is a local $\mc{C}^1$ diffeomorphism on  $\wh{\mc{D}}_{\vsg}$ such that, for any $(u,v) \in \wh{\mc{D}}_{\vsg}$, there exist $N$-independent $s, s^{\prime}>0$
such that
\beq
\bs{\Phi}\; : \; B_{(u,v),s} \tend \bs{\Phi} \Big( B_{(u,v),s} \Big) \supset B_{\bs{\Phi}(u,v),s^{\prime}}
\enq
is a diffeomorphism, with $B_{a,r}$ being the open ball of radius $r$ centred at $a$.

This is enough so as to ensure that $\bs{\Phi}$ is a $\mc{C}^1$ diffeomorphism on $\wh{\mc{D}}_{\vsg}$. Indeed, assume that there exist $\big( u, v \big), \big(u^{\prime}, v^{\prime}\big) \in \wh{\mc{D}}_{\vsg}$,
$\big( u, v \big) \not= \big(u^{\prime}, v^{\prime}\big)$ such that $\bs{\Phi}\big( u, v \big)\, = \,  \bs{\Phi}\big(u^{\prime}, v^{\prime}\big)$. Then, one has the relation
\beq
\left\{ \ba{ccccc}   u^{\prime}-u & = &  \de \Phi_1\big( u, v\big) \, - \,  \de \Phi_1\big( u^{\prime}, v^{\prime} \big)  & = & \e{O}\big( N^{ - \vsg \wt{\varkappa}_{\eta }(1-\eta^{\prime}) } \big) \vspace{2mm}  \\
v^{\prime}-v & = &  \de \Phi_2\big( u, v\big) \, - \,  \de \Phi_2\big( u^{\prime}, v^{\prime} \big) & = & \e{O}\big( N^{ - \vsg \wt{\varkappa}_{\eta }(1-\eta^{\prime}) } \big) \ea \right.  \;.
\enq
However, $\bs{\Phi}$ is injective on $B_{(u,v),s} \ni (u^{\prime}, v^{\prime})$ what entails $\big( u, v \big)  = \big(u^{\prime}, v^{\prime}\big)$, a contradiction.
This entail that $\bs{\Phi}$ is a diffeomorphism on $\wh{\mc{D}}_{\vsg}$.

Finally, by the estimates on $\de \bs{\Phi}$, it holds that $\bs{\Phi}\Big( \mf{c}_0 \, , \, \tfrac{ \a \mf{c}_0 }{ \pi (\om_1 + \om_2) }  \Big)
\, = \, \e{O}\big( N^{ - \vsg \wt{\varkappa}_{\eta }(1-\eta^{\prime}) } \big) $. However, since there exists $s, s^{\prime}>0$
such that
\beq
\bs{\Phi} \Big( B_{ \big( \mf{c}_0 \, , \, \tfrac{ \a \mf{c}_0 }{ \pi (\om_1 + \om_2) }  \big) ,s} \Big) \supset B_{\bs{\Phi} \big( \mf{c}_0 \, , \, \tfrac{ \a \mf{c}_0 }{ \pi (\om_1 + \om_2) }  \big),s^{\prime}} \;,
\enq
it follows that $(0,0) \in \bs{\Phi} \big(  \wh{\mc{D}}_{\vsg} \big)$, what ensures the existence and uniqueness of solutions to the
system \eqref{ecriture contraintes sur pts bord aN et bN} on $\mc{D}_{\vsg}$.

The form of the leading large-$N$ behaviour for $a_{N;\a}, b_{N;\a}$ then follows readily. \qed

\begin{lemme}
\label{Lemma DA des constantes definissant bN et aN}
 The following large-$N$ asymptotics hold
\beq
\ex{ \ov{b}_{N;\a} } \; = \; \f{ \mf{c}_0 N }{ 2 } \, + \, \f{ \a \mf{c}_0 }{ 2\pi (\om_1+\om_2) } \, - \,
 \f{2\i R_{\ua}(\i) }{ N \mf{c}_0 R_{\da}(\i)  } \bigg(  1 \, + \, \f{2 \mc{F}[\op{g}](\i) }{ \pi r } \bs{1}_{1<\zeta}  \bigg) \, + \, \e{O}\Big( \f{1}{N^{\varkappa_{\eta}} }\Big)
\label{ecriture DA exp bN alpha}
\enq
and
\beq
\ex{ -\ov{a}_{N;\a} } \; = \; \f{ \mf{c}_0 N }{ 2 } \, - \, \f{ \a \mf{c}_0 }{ 2\pi (\om_1+\om_2) } \, - \,
 \f{2\i R_{\ua}(\i) }{ N \mf{c}_0 R_{\da}(\i)  } \bigg(  1 \, + \, \f{2 \mc{F}[\op{g}](\i) }{ \pi r } \bs{1}_{1<\zeta}   \bigg) \, + \, \e{O}\Big( \f{1}{N^{\varkappa_{\eta}} }\Big)
\label{ecriture DA exp moins aN alpha}
\enq
These asymptotic expansions involve the constant
\beq
\mf{c}_0 \, = \, \f{4\pi }{ \mf{r} } \sqrt{\om_1+\om_2} R_{\ua}(\i)
\enq
while $\varkappa_{\eta}$ is as introduced in \eqref{definition varkappa eta et zeta}. Moreover, it also holds that
\bem
\ov{b}_{N;\a} \,  = \,  \ln \bigg( \f{ \mf{c}_0 N }{ 2 } \bigg) \, + \, \f{ \a }{ \pi N (\om_1+\om_2) }  \\ \, - \,
\f{1}{N^2} \bigg\{  \f{4\i R_{\ua}(\i) }{ \mf{c}_0^2 R_{\da}(\i)  } \bigg(  1 \, + \, \f{2 \mc{F}[g](\i) }{ \pi \mf{r} } \bs{1}_{1<\zeta}   \bigg)
\, + \,  \f{\a^2 }{2\pi^2   (\om_1+\om_2)^2} \bigg\} \, + \, \e{O}\Big( \f{1}{N^{\varkappa_{\eta}+1} }\Big)
\end{multline}
and
\bem
\ov{a}_{N;\a} \,  = \,  -\ln \bigg( \f{ \mf{c}_0 N }{ 2 } \bigg) \, + \, \f{ \a }{ \pi N (\om_1+\om_2) }  \\
\, + \, \f{1}{N^2} \bigg\{  \f{4\i R_{\ua}(\i) }{ \mf{c}_0^2 R_{\da}(\i)  } \bigg(  1 \, + \, \f{2 \mc{F}[\op{g}](\i) }{ \pi \mf{r} } \bs{1}_{1<\zeta}   \bigg)
\, - \,  \f{\a^2 }{2\pi^2   (\om_1+\om_2)^2} \bigg\} \, + \, \e{O}\Big( \f{1}{N^{\varkappa_{\eta}+1} }\Big)
\end{multline}

\end{lemme}

\Proof

Starting from the expressions of Propositions \ref{Proposition contrainte cpte en aN et bN explicitee en DA} and \ref{Proposition calcul points de bord aN et bN}, one infers that
\beq
\ex{\ov{b}_{N} } \, -\,\ex{-\ov{a}_{N} }
 \; = \; 2 \f{  \a  \chi_{11;-}(0) }{ \mf{r}   \chi_{11}(i) }  \cdot \bigg\{  1 \, + \, \f{2 \i \mc{F}[\op{g}](\i) \, \ex{-\ov{x}_N} }{ \pi \mf{r} \chi_{11}(\i) R_{\ua}(-\i) } \bs{1}_{1<\zeta}  \bigg\}^{-1}
 \; + \; \e{O}\Big(    \ex{- \ov{b}_N \varkappa_{\eta}  } + \ex{  \ov{a}_N \varkappa_{\eta}  }  \Big)  \;,
\enq
and
\bem
 \ex{\ov{b}_{N} } \, + \,\ex{-\ov{a}_{N} }  =   \f{ 4 \pi N }{ \i \, \mf{r}   } \cdot
\f{  1   +   \f{ \a }{2\pi N}  \Big[ \chi_{11;-}^{\prime}(0)\chi_{12;-}(0)  -   \chi_{12;-}^{\prime}(0)\chi_{11;-}(0)
  +   \tfrac{1}{2}  \chi_{11;-}^2(0)  \cdot \bigg\{  1   +  \tfrac{2 \i \mc{F}[\op{g}](\i) \ex{-\ov{x}_N} }{ \pi \mf{r} \chi_{11}(\i) R_{\ua}(-\i) } \bs{1}_{1<\zeta}  \bigg\}^{-1} \Big] }
{ \chi_{11;-}(0) \chi_{12}(\i) \,- \, \chi_{12;-}(0) \chi_{11}(\i) \, - \, \tfrac{\i}{2}  \chi_{11;-}(0) \chi_{11}(\i)
-  \tfrac{2 \i \mc{F}[\op{g}](\i) \ex{-\ov{x}_N} }{ \pi \mf{r} R_{\da}(\i)  R_{\da}(0) } \bs{1}_{1<\zeta}  }  \\
\; + \;  \e{O}\Big(    \ex{- \ov{b}_N \varkappa_{\eta}  } + \ex{  \ov{a}_N \varkappa_{\eta}  }  \Big) \;.
\end{multline}
Then, by using the asymptotic expansions given in \eqref{ecriture DA dominant chi dans region sous R} and \eqref{ecriture DA dominant chi dans region entre R + i eps et Gamma up}
one gets that
\bem
 \ex{\ov{b}_{N} } + \ex{-\ov{a}_{N} }  = N \mf{c}_0
 \f{  1- \tfrac{ \a }{ \pi N} \cdot \tfrac{ \i R_{\ua}(\i) }{ R_{\da}(\i) } \cdot   \tfrac{ 2  \mc{F}[\op{g}](\i)  }{ \pi \mf{r} (\om_1+ \om_2)} \ex{-\ov{x}_N} \bs{1}_{1<\zeta}
\, + \, \tfrac{1}{N} \e{O}\Big(  \a \ex{-\zeta(1-\eta)\ov{x}_N }   \,+\,\a \ex{-\ov{x}_N}  \bs{1}_{1<\zeta}  \Big)  }
{ 1+   \tfrac{ \i R_{\ua}(\i) }{ R_{\da}(\i) } \ex{-\ov{x}_N}  \Big[ 1+ \tfrac{2  \mc{F}[\op{g}](\i)  }{ \pi \mf{r}  } \bs{1}_{1<\zeta}  \Big] \, + \, \e{O}\Big(  \a \ex{-\zeta(1-\eta)\ov{x}_N }  \Big) } \\
\;+ \; \e{O}\Big(    \ex{- \ov{b}_N \varkappa_{\eta}  } + \ex{  \ov{a}_N \varkappa_{\eta}  }  \Big) \;,
\end{multline}
and
\beq
\ex{\ov{b}_{N} } \, -\,\ex{-\ov{a}_{N} } \; = \; \f{ \a \mf{c}_0}{ \pi (\om_1+\om_2)  } \; + \;  \e{O}\Big(    \ex{- \ov{b}_N \varkappa_{\eta}  } + \ex{  \ov{a}_N \varkappa_{\eta}  }  \Big) \;.
\enq
From there on, the result follow from straightforward algebra.  \qed

\section{Large-$N$ behaviour of the interpolating integral}
\label{Section DA integrale interpolante}

\begin{prop}
\label{Proposition interpolating integral}

 It holds
\beq
\Int{ a_{N;\a} }{ b_{N;\a} } \hspace{-1mm}  \dd \xi \, \xi\, \wh{\varrho}_{\e{eq};\a}(\xi) \; = \;
\f{ \a }{ \pi (\om_1+\om_2) N \tau_N } \bigg[ \ln \Big( \tfrac{N \mf{c}_0}{2} \Big) \, + \, \i \ln^{\prime}\! R_{\da}(0) \bigg]
 \, + \, \e{O}\bigg( \f{\tau_N}{N} \Big[  N^{-1} \bs{1}_{1<\zeta}  \, +  \,  N^{ - \varkappa_{\eta} } \Big] \bigg)
\enq

\end{prop}

\Proof

One starts by observing that
\beq
\Int{ a_{N;\a} }{b_{N;\a}} \hspace{-1mm}   \dd \xi \ex{-\i \la \tau_N (\xi-a_{N;\a})} \xi \; = \;    \f{a_{N;\a} \, - \, b_{N;\a} \ex{-\i \la \ov{x}_{N;\a} } }{ \i \la \tau_N }
\, + \,  \f{ \ex{-\i \la \ov{x}_{N;\a} }\, - \, 1  }{ (\la \tau_N)^2 }\;.
\enq
Here, we agree that $ \ov{x}_{N;\a}\, = \, \ov{b}_{N;\a}\, - \, \ov{a}_{N;\a}$.
Thus, according to the partially integrated expression for $\mc{W}_N\big[ V_{N;\a}^{\prime} \big]$
obtained in Proposition \ref{ecriture expression exacte pour inverse sur VN alpha prime}, one may decompose the integral into three terms
\beq
\Int{ a_{N;\a} }{ b_{N;\a} } \hspace{-1mm} \dd \xi \, \xi \, \wh{\varrho}_{\e{eq};\a}(\xi) \; = \; \sul{a=1}{3} \mc{H}^{(a)}
\enq
where
\beq
\mc{H}^{(1)} \ = \, \f{ \mf{r} \tau_N }{ 4\i\pi N  } \hspace{-2mm} \Int{ \R + 2\i\eps^{\prime} }{} \hspace{-2mm} \f{ \dd \la }{ 2\i\pi }
\bigg\{ \f{ \chi_{12}(\la) }{ \la } \msc{U}_{12}(\la) \; + \; \chi_{11}(\la) \msc{U}_{11}(\la) \bigg\}
\cdot \bigg\{  \f{a_{N;\a} \, - \, b_{N;\a} \ex{-\i \la \ov{x}_{N;\a} } }{ \i \la \tau_N } \, + \,  \f{ \ex{-\i \la \ov{x}_{N;\a} }\, - \, 1  }{ (\la \tau_N)^2 } \bigg\} \;,
\enq
\beq
\mc{H}^{(2)} \ = \, \f{ \a \tau_N }{ 2 \i \pi N  } \hspace{-2mm} \Int{ \R + 2\i\eps^{\prime} }{} \hspace{-2mm} \f{ \dd \la }{ 2\i\pi }
 \chi_{11}(\la) \chi_{12;-}(0)
\cdot \bigg\{  \f{a_{N;\a} \, - \, b_{N;\a} \ex{-\i \la \ov{x}_{N;\a} } }{ \i \la^2 \tau_N } \, + \,  \f{ \ex{-\i \la \ov{x}_{N;\a} }\, - \, 1  }{ \la^3 \tau_N^2 } \bigg\} \;,
\enq
and, by employing the notations introduced in Proposition \ref{ecriture expression exacte pour inverse sur VN alpha prime}
\beq
\mc{H}^{(3)} \ = \, \mc{H}^{(3)}_{\ua} \, +\,  \mc{H}^{(3)}_{\da} \, + \, \mc{H}^{(3)}_{0} \qquad \e{with} \qquad
\mc{H}^{(3)}_{\ups}\,= \,  \Int{ a_{N;\a} }{ b_{N;\a} } \hspace{-2mm} \dd \xi \, \xi \varpi_{N;\ups}^{(3)} (\xi)  \;, \quad \ups \in \{ \ua, \da, 0\}\;.
\enq
We stress that now, all the above quantities involve the solution $\chi$ subordinate to the choice of endpoints $a_{N;\a} , \, b_{N;\a}$.

The first two contributions can be computed in closed form. Indeed, by applying the previously introduced notations, one gets
\beq
\mc{H}^{(2)} \ = \, \f{ \a \tau_N }{ 2 \i \pi N  }  \chi_{12;-}(0) \hspace{-2mm} \Int{ \R + 2\i\eps^{\prime} }{} \hspace{-2mm} \f{ \dd \la }{ 2\i\pi }
\cdot \bigg\{  \f{a_{N;\a}  \chi_{11}(\la) \, - \, b_{N;\a}  \chi_{11;-}(\la) }{ \i \la^2 \tau_N } \, + \,  \f{ \chi_{11;-}(\la)\, - \, \chi_{11}(\la)  }{ \la^3 \tau_N^2 } \bigg\}   \;.
\enq
Due to the $\e{O}\big( \la^{-\tf{5}{2}} \big)$ decay of the integrand at $\infty$
\begin{itemize}

 \item[i)] the contribution of the integrand involving  $\chi_{11}$
can be evaluated by taking the residues of the integrand's poles located above $\R + 2\i\eps^{\prime}$. Since there are no poles, this part produces $0$.

\item[ii)] The contribution of the integrand involving  $\chi_{11;-}$
can be evaluated by taking the residues of the integrand's poles located below $\R + 2\i\eps^{\prime}$. The only poles present are the third and second order poles at $0$.

\end{itemize}
All-in-all, one gets that
\beq
\mc{H}^{(2)} \ = \, - \f{ \a \tau_N }{ 2 \i \pi N  }  \chi_{12;-}(0)
\cdot \bigg\{  - \f{   b_{N;\a}   }{ \i  \tau_N }  \chi_{11;-}^{\prime}(0) \, + \,  \f{ 1  }{ 2 \tau_N^2 } \chi_{11;-}^{\prime\prime}(0) \bigg\}   \;.
\enq
The large-$N$ behaviour of $\chi_{1a}$ given in \eqref{ecriture decomposition chi fct dominante et perturbative sour R}, allows one to infer that
\beq
\chi_{12;-}(0) \; = \; \e{O}\Big( \ex{-\zeta (1-\eta) \ov{x}_{N;\a}} \Big) \quad \e{and} \quad
 \chi_{11;-}^{(k)}(0)\; = \; \e{O}\Big(  \ov{x}_{N;\a}^{k} \Big) \quad \e{with} \quad k \in \mathbb{N}.
\enq
This entails that
\beq
\mc{H}^{(2)} \, = \,     \e{O}\bigg(  \f{  |\a|  \tau_N }{   N  } \cdot \ex{-\zeta (1-\eta) \ov{x}_{N;\a}} \bigg)
\, = \,   \e{O}\bigg(  \f{  |\a|  \tau_N }{   N^{1+2\zeta (1-\eta)}  } \bigg)  \;.
\enq

Similar handlings lead to
\bem
\mc{H}^{(1)} \ = \, - \f{ \mf{r} \tau_N }{ 4\i\pi N  }\Bigg\{   -  \f{ b_{N;\a} }{ \i \tau_N }  \Dp{\la}\Big( \chi_{12;-}\msc{U}_{12}\Big)(0)
\, + \,  \f{ 1 }{2 \tau_N^2}  \Dp{\la}^2\Big(\chi_{12;-} \msc{U}_{12}\Big)(0)  \\
\, - \,  \f{ b_{N;\a} }{ \i \tau_N } \Big(\chi_{11} \msc{U}_{11}\Big)(0) \, + \,  \f{ 1 }{ \tau_N^2 }  \Dp{\la}\Big(\chi_{11;-} \msc{U}_{11}\Big)(0)  \Bigg\} \;.
\end{multline}
A long but straightforward calculation utilising the expansion  \eqref{ecriture decomposition chi fct dominante et perturbative sour R} yields
\bem
\mc{H}^{(1)} \ = \,   \f{ \mf{r} \tau_N }{ 4\i\pi N  R_{\da}(0) R_{\ua}(\i) }\Bigg\{  \ov{b}_{N;\a} \ex{\ov{b}_{N;\a}}
\, + \,  \ov{a}_{N;\a} \ex{-\ov{a}_{N;\a}} \, - \, \big( \ex{\ov{b}_{N;\a}}  -  \ex{-\ov{a}_{N;\a}} \big)
\cdot \big(1-\i \ln^{\prime}\! R_{\da}(0) \big) \\
\, + \, \i \f{ R_{\ua}(\i) }{ R_{\da}(\i) } \cdot \Big[ \ov{b}_{N;\a} \ex{-\ov{b}_{N;\a}} \, + \,  \ov{a}_{N;\a} \ex{\ov{a}_{N;\a}} \, - \, \ex{-\ov{x}_{N;\a}}  \big( \ex{\ov{b}_{N;\a}}  -  \ex{-\ov{a}_{N;\a}} \big)
\cdot \big(1+\i \ln^{\prime}\! R_{\da}(0) \big)  \Big] \Bigg\} \Big( 1+\e{O}\big( \tau_N \ex{-\zeta (1-\eta) \ov{x}_{N;\a} } \big) \Big) \;.
\end{multline}
Observe that, for large N, one has
\beq
\ov{b}_{N;\a} \ex{\ov{b}_{N;\a}} \, + \,  \ov{a}_{N;\a}\ex{-\ov{a}_{N;\a}} \, = \, v_{N;\a} \ln \bigg( \f{N u_{N;\a} \ex{} }{ 2 } \bigg)  \, + \, \e{O}\big( N^{-2} \big)
\enq
and
\beq
\ov{b}_{N;\a} \ex{-\ov{b}_{N;\a}} \, + \,  \ov{a}_{N;\a} \ex{\ov{a}_{N;\a}}\, = \, \e{O}\bigg(  \f{ v_{N;\a} \tau_N   }{ (Nu_{N;\a})^2 }\bigg) \;.
\enq
Finally, by inserting the large-$N$ expansion of the endpoints obtained in Lemma \ref{Lemma DA des constantes definissant bN et aN}, one gets
\beq
 \mc{H}^{(1)}  \ = \,  \f{ \a }{ \pi (\om_1+\om_2) N \tau_N } \bigg[ \ln \Big( \tfrac{N \mf{c}_0}{2} \Big) \, + \, \i \ln^{\prime}\! R_{\da}(0) \bigg]
\, - \, \f{ 2 \a^2 }{ \pi^2 (\om_1+\om_2)^2 N^2 \tau_N } \, + \, \e{O}\bigg( \f{1}{N^3} \, +  \, \f{\ex{-\zeta (1-\eta) \ov{x}_{N;\a}} }{ N } \bigg) \;.
\enq

Hence, to conclude, it remains to estimate $\mc{H}^{(3)}$. We estimate separately each of the $\mc{H}^{(3)}_{\ups}$.
For $\ups \in \{\ua, \da\}$, by taking the expressions for $\varpi_{N;\ua/\da}^{(3)}$ obtained in Proposition \ref{ecriture expression exacte pour inverse sur VN alpha prime}
one gets the representation
\beq
\mc{H}_{\ups}^{(3)} \, = \, \f{  \eps_{\ups}   }{ 4\i\pi N \tau_N } \hspace{-2mm} \Int{ \R + 2\i\eps^{\prime} }{} \hspace{-2mm} \f{ \dd \la }{ 2 \i \pi \la^{3} }
\Big( \bs{F}_{\ua}(\la)\, + \, \ex{-\i\la \ov{x}_N}  \bs{F}_{\da}(\la), \bs{\mc{E}}^{(\ups)}(\la) \Big) \;,
\enq
where we have introduced
\beqa
\bs{F}_{\da}(\la) & = & \big( \i \ov{b}_{N;\a} \la + 1 \big)  \bs{E}_L^{(\ua)}(\la)  \\
\bs{F}_{\ua}(\la) & = & -\,\big( \i \ov{a}_{N;\a} \la + 1 \big)  \bs{E}_L^{(\ua)}(\la) \,-\,  \big( \i \ov{b}_{N;\a} \la + 1 \big)  \bs{E}_L^{(\da)}(\la)
\, +\, \ex{\i \la \ov{x}_{N;\a} } \big( \i \ov{a}_{N;\a} \la + 1 \big)  \bs{E}_L^{(\ua)}(\la) \;.
\eeqa
Thus, upon introducing $\wt{\varkappa}_{\eta}\, = \, (1-\eta) \e{min}\big\{  1, \zeta \big\}$ and deforming the integration contours, one gets
\bem
\mc{H}_{\ups}^{(3)} \, = \, \f{  \eps_{\ups}   }{ 4\i\pi N \tau_N } \hspace{-2mm} \Int{ \R +  \i \wt{\varkappa}_{\eta} }{} \hspace{-2mm} \f{ \dd \la }{ 2 \i \pi \la^{3} }
\Big( \bs{F}_{\ua}(\la), \bs{\mc{E}}^{(\ups)}(\la) \Big) \, -  \, \f{  \eps_{\ups}   }{ 8\i\pi N \tau_N } \Dp{\la}^2 \bigg\{   \ex{-\i\la \ov{x}_N }   \Big( \bs{F}_{\da}(\la) , \bs{\mc{E}}^{(\ups)}(\la) \Big)   \bigg\}_{\mid \la=0}   \\
 \, + \, \f{  \eps_{\ups}   }{ 4\i\pi N \tau_N } \hspace{-2mm} \Int{ \R - \i \wt{\varkappa}_{\eta} }{} \hspace{-2mm} \f{ \dd \la }{ 2 \i \pi \la^{3} }
  \ex{-\i\la \ov{x}_N }  \Big( \bs{F}_{\da}(\la), \bs{\mc{E}}^{(\ups)}(\la) \Big)
\end{multline}
Then, using that within the band $|\Im(\la)| \leq \wt{\varkappa}_{\eta}$ one has the bounds
\beq
 \norm{  \bs{\mc{E}}^{(\ups)}(\la)  } \, \leq \, \f{ C }{ 1+|\la|  } \Big( N^{-1} \bs{1}_{1<\zeta} \, + \, N^{-\varkappa_{\eta}} \Big) \;,
\enq
one readily gets that
\beq
\Big| \mc{H}_{\ups}^{(3)}  \Big| \, \leq  \, \f{ C \tau_N}{ N }\Big( N^{-1} \bs{1}_{1<\zeta} \, + \, N^{-\varkappa_{\eta}} \Big) \;.
\enq
Finally, we focus on $ \mc{H}_{0}^{(3)}$ which, upon using that $\Big(  \bs{E}_L^{(\da)}(\la),  \bs{E}_R^{(\da)}(\la) \Big)=0$,  may be recast as
\beq
 \mc{H}_{0}^{(3)}\, =  \,  \f{    1 }{ 4\i\pi N \tau_N  } \hspace{-2mm} \Int{ \R + 2\i\eps^{\prime} }{} \hspace{-2mm} \f{ \dd \la }{ 2\i\pi \la^2 }
\f{  \mc{F}[\op{g}](\la)  }{  \cosh\Big[ \tfrac{\pi \la }{2} \Big] }  \Big( \bs{E}_L^{(\ua)}(\la), \bs{E}^{(\da)}_R(\la) \Big)
\Big\{ \big(\i \ov{a}_{N;\a} \la + 1 \big) \ex{ - \i \la \ov{a}_{N;\a} }  \,- \, \ex{ - \i \la \ov{b}_{N;\a} }   \big( \i \ov{b}_{N;\a} \la + 1 \big)   \Big\} \;.
\enq
Then, it is enough to observe that throughout the strip $|\Im(\la)|\leq \wt{\varkappa}_{\eta}$
\beq
 \Big( \bs{E}_L^{(\ua)}(\la), \bs{E}^{(\da)}_R(\la) \Big) \; = \; - \f{ \la }{ R(\la) } \, + \, \mf{R}_N(\la) \qquad \e{with} \qquad
  \mf{R}_N(\la) \, = \, \e{O}\Big( (1+|\la|) \ex{-\zeta  (1-\eta) \ov{x}_{N;\a} } \Big) \;.
\enq
In particular, the leading term has a second order zero at $\la=0$. This leads to
\bem
 \mc{H}_{0}^{(3)}\, =     \hspace{-2mm} \Int{ \R +  \i\wt{\varkappa}_{\eta}  }{} \hspace{-2mm} \f{ \dd \la }{ 2\i\pi \la^2 }
\f{  \mc{F}[\op{g}](\la)  }{  \cosh\Big[ \tfrac{\pi \la }{2} \Big] }  \Big( \bs{E}_L^{(\ua)}(\la), \bs{E}^{(\da)}_R(\la) \Big)  \f{   \i \ov{a}_{N;\a} \la + 1   }{ 4\i\pi N \tau_N  }  \ex{ - \i \la \ov{a}_{N;\a} } \\
\, -   \hspace{-2mm} \Int{ \R -  \i\wt{\varkappa}_{\eta}  }{} \hspace{-2mm} \f{ \dd \la }{ 2\i\pi \la  }
\f{  \mc{F}[\op{g}](\la)  }{  \cosh\Big[ \tfrac{\pi \la }{2} \Big]  R(\la) }   \f{    \i \ov{b}_{N;\a} \la + 1  }{ 4\i\pi N \tau_N  } \ex{ - \i \la \ov{b}_{N;\a} }
\, +   \hspace{-2mm} \Int{ \R +  2\i \eps^{\prime}   }{} \hspace{-2mm} \f{ \dd \la }{ 2\i\pi \la^2 }
\f{  \mc{F}[\op{g}](\la)  }{  \cosh\Big[ \tfrac{\pi \la }{2} \Big]    }   \mf{R}_N(\la)   \f{    \i \ov{b}_{N;\a} \la + 1  }{ 4\i\pi N \tau_N  }  \ex{ - \i \la \ov{b}_{N;\a} } \;.
\end{multline}
Upon deforming further the integrals up to $ \R  \pm  \i\varkappa_{\eta}$ and picking the residues of the simple pole at $\pm \i $ and then  applying
direct bounds, one eventually gets
\beq
\Big|  \mc{H}_{0}^{(3)} \big| \, \leq  \,  \f{ C }{ N \tau_N } \bigg\{  \tau_N\Big[ N^{-1} \bs{1}_{1<\zeta} + N^{-\varkappa_{\eta}} \Big] \, + \, \tau_N
\ex{-\zeta (1-\eta) \ov{x}_{N;\a} + 2\eps^{\prime}\ov{b}_{N;\a} }  \bigg\} \; \leq \; \f{C}{N} \Big[ N^{-1} \bs{1}_{1<\zeta} + N^{-\varkappa_{\eta}} \Big] \;.
\enq
The claim then follows by putting the various estimates together. \qed

\section*{Conclusion}

In this work, we have provided a full characterisation of the equilibrium measure which governs the leading asymptotic expansion of the
logarithm of the Lukyanov integral. This allowed us to check, by means of explicit calculations the predictions relative to the
leading term of the Lukyanov conjecture describing the large-$N$ behaviour of a multiple-integral supposed to
provide the lattice regularisation of the vacuum expectation value of the  exponential of the field operator in the quantum Sinh-Gordon finite volume $R$
field theory. Our calculations confirm this part of the conjecture. However, a lack of sharp bounds on the remainder, issuing from our incapacity to control,
on sufficiently fine scales, the inverse of the master operator arising in the system of loop equations, does not allow us to prove that indeed the other corrections
which could contribute to the asymptotics of the derivative $\Dp{\a} \ln \mc{Z}_N\big[ V_{N;\a}\big]$ will not do so on a stronger than $\ln N$ scale.
It would be extremely interesting to develop a much better understanding of the scaling regimes of the master operator appropriate for this setting.

We plan to address these questions, in full rigour, by alternative methods in further works.

\section*{Acknowledgements}
 KKK is supported by the
CNRS. CDG and KKK are supported  by the ERC Project LDRAM: ERC-2019-ADG Project 884584 and by the
CNRS 80Prime Grant "Asymptotiques d'intégrales multiples associées à la séparation des variables quantiques".
The authors thank Alice Guionnet  for numerous stimulating discussions
related to the topics tackled in this paper.

\end{document}